\documentclass[times, twocolumn, tighten]{aastex62}
\usepackage{newtxmath}
\usepackage[USenglish]{babel}
\usepackage{amsmath,amsfonts}		% Bold symbol
\usepackage{natbib}
\usepackage{tensor}			% Nuclide symbol

\usepackage[artemisia]{textgreek}	% Upright Greek letter in text mode,
\usepackage{bookmark}

\newcommand{\vv}[1]{\boldsymbol{#1}}		% vector
\newcommand{\logg}{\log_{10}}			% logarithm base-10
\newcommand{\rd}{r_\text{d}}			% sound scale at drag epoch
\newcommand{\lcdm}{\textLambda{}CDM }		% Lambda-CDM
\newcommand{\densb}{\Omega_{\text{b}}}	% baryon fraction
\newcommand{\densm}{\Omega_{\text{m}}}	% matter fraction
\newcommand{\fb}{f_{\text{b}}}		% baryon to total matter ratio
\newcommand{\dA}{d_{\text{A}}}		% angular diameter dist.
\newcommand{\dL}{d_{\text{L}}}		% luminosity dist.
\newcommand{\muL}{\mu_{\text{L}}}		% luminosity dist. modulus
\newcommand{\mulc}{\mu_{\text{Lc}}}		% ditto, compressed
\newcommand{\eref}[1]{Equation~(\ref{#1})}	% cross-linking equations
\newcommand{\erefs}[1]{Equations~(\ref{#1})}	% same as \eref, with plural

\defcitealias{2016MNRAS.463.1651M}{M16}		% M16 paper, frequently cited.

\begin{document}

\shorttitle{Distance--Duality Tests with SNIa and BAO}
\shortauthors{Ma \& Corasaniti}

\title{Statistical Test of Distance--Duality Relation with Type Ia Supernovae
and Baryon Acoustic Oscillations}
\received{2018 February 12}
\revised{2018 April 18}
\accepted{2018 May 25}
\published{2018 July 12}
\submitjournal{\apj}

\author[0000-0001-6109-2365]{Cong Ma}
\affiliation{Purple Mountain Observatory, Chinese Academy of Sciences, 8
Yuanhua Road, 210034 Nanjing, People's Republic of China}
\affiliation{Graduate School, University of the Chinese Academy of Sciences,
19A Yuquan Road, 100049 Beijing, People's Republic of China}

\author[0000-0002-6386-7846]{Pier-Stefano Corasaniti}
\affiliation{LUTH, UMR 8102 CNRS, Observatoire de Paris, PSL Research
University, Universit{\'e} Paris Diderot, \\
5 Place Jules Janssen, 92190 Meudon, France}

\correspondingauthor{Cong Ma}
\email{cma@pmo.ac.cn}

\begin{abstract}
    We test the distance--duality relation $\eta \equiv \dL / [ (1 + z)^2 \dA
    ] = 1$ between cosmological luminosity distance ($\dL$) from the JLA SNe
    Ia compilation and angular-diameter distance ($\dA$) based on Baryon
    Oscillation Spectroscopic Survey (BOSS) and WiggleZ baryon acoustic
    oscillation measurements.  The $\dL$ measurements are matched to $\dA$
    redshift by a statistically consistent compression procedure.  With Monte
    Carlo methods, nontrivial and correlated distributions of $\eta$ can be
    explored in a straightforward manner without resorting to a particular
    evolution template $\eta(z)$.  Assuming independent constraints on
    cosmological parameters that are necessary to obtain $\dL$ and $\dA$
    values, we find 9\% constraints consistent with $\eta = 1$ from the
    analysis of SNIa + BOSS and an 18\% bound results from SNIa + WiggleZ.
    These results are contrary to previous claims that $\eta < 1$ has been
    found close to or above the $1 \sigma$ level.  We discuss the effect of
    different cosmological parameter inputs and the use of the apparent
    deviation from distance--duality as a proxy of systematic effects on
    cosmic distance measurements.  The results suggest possible systematic
    overestimation of SNIa luminosity distances compared with $\dA$ data when
    a \textit{Planck} \lcdm cosmological parameter inference is used to
    enhance the precision.  If interpreted as an extinction correction due to
    a gray dust component, the effect is broadly consistent with independent
    observational constraints.
\end{abstract}


\section{Introduction}
\label{sec:intro}

A generic property of cosmological distances in general relativity is that the
angular-diameter distance $\dA$ and the luminosity distance $\dL$ to the same
cosmological redshift $z$ satisfy the distance--duality (DD) relation
\citep{E71, 2009GReGr..41..581E}
\begin{equation}
    \eta \equiv \frac{\dL(z)}{(1 + z)^2 \dA(z)} = 1.
    \label{eq:dd}
\end{equation}
The theoretical underpinnings of this relation are the geometrical reciprocity
relation \citep{E33, 2007GReGr..39.1055E} that holds in any metric theory of
gravity and the fact that light propagates along null geodesics with the
photon number conserved.  This is not the case in nonmetric theories of
gravity, theories of varying fundamental constants, or axion--photon mixing
models \citep{2004PhRvD..69j1305B, 2004PhRvD..70h3533U}.  Therefore, an
observational falsification of \eref{eq:dd} could be a useful probe of exotic
physics, provided that cosmic distance measurements are exempt from
astrophysical systematic effects \citep[e.g., see][for violations induced by
intergalactic dust extinction]{2006MNRAS.372..191C}.

Several cosmological probes can be used for this purpose.  Luminosity
distances are indicated by the brightness of Type Ia supernovae (SNIa) through
the standard-candle relation, and angular-diameter distances can be inferred
from the apparent size of cosmic standard rulers. Numerous studies devoted to
testing the validity of \eref{eq:dd} have used SNIa $\dL$ data in combination
with $\dA$ estimates from X-ray observations of galaxy clusters
\citep[e.g.,][]{2004PhRvD..69j1305B, 2004PhRvD..70h3533U, 2006IJMPD..15..759D,
2010ApJ...722L.233H, 2015JCAP...10..061S}. However, angular distance
measurements from galaxy cluster observations are cosmological-model dependent
\citep[see, e.g.,][]{2006ApJ...647...25B}.  Furthermore, astrophysical
uncertainties such as the three-dimensional profile of intra-cluster plasma
may significantly affect the estimation of $\dA$ \citep{2012ApJ...745...98M,
2013ApJ...777L..24Y}. Recent works have instead used angular-diameter distance
measurements from observations of strong lens systems
\citep{2016JCAP...02..054H, 2017JCAP...09..039H, 2016ApJ...822...74L,
2017IJMPD..2650097F, 2017JCAP...07..010R}. These estimates have the advantage
of being cosmological-model independent, but they are not exempt from
systematic effects. In fact, lens mass model uncertainties due to the
mass-sheet degeneracy and the effect of external perturbators may have a
strong impact on the inferred properties of these systems
\citep{2013AA...559A..37S}.  More reliable estimates can be derived from the
baryon acoustic oscillation (BAO) signal in the galaxy power spectrum.
Although they depend on the cosmic matter density and the Hubble constant,
thus demanding prior external information, these only contribute to
statistical uncertainties.  In contrast, systematic effects due to the
nonlinearity of the matter density field are largely sub-dominant, as they
are expected to alter $\dA$ estimates to less than a few percent level
\citep{2008PhRvD..77b3533C, 2014MNRAS.440.1420R}.

A common approach to test the DD relation uses distance measurements to
constrain parameterizations of $\eta$ as a function of redshift $z$
\citep[e.g.][]{2012JCAP...12..028N, 2015PhRvD..92b3520W, 2016JCAP...02..054H,
2017JCAP...09..039H, 2016ApJ...822...74L, 2017IJMPD..2650097F}.  The choice of
an $\eta(z)$ template function imposes a strong prior on the DD analysis,
which may result in different outcomes depending on its form. In our view, a
template-free study is preferred because of its generality and robustness in
the absence of abundant data.

A practical issue underlying the tests is the fact that measurements of $\dL$
and $\dA$ may not be available at the same redshift.  Several studies have
attempted to address this problem by selecting data points under a proximity
criterion, e.g., by only using data within a redshift separation $\left\vert
\Delta z \right\vert \le 5 \times 10^{-3}$ \citep{2010ApJ...722L.233H}.
However, this may incur the penalty of significantly reduced statistical
information encoded in the data sets.  Moreover, it does not guarantee that the
data points thus chosen provide a representative local sample.  This situation
is analogous to the problem of estimating the cross-correlation of unevenly
sampled time-series data, for which narrow windows centered on cherry-picked
data can lead to a spuriously high significance of detection
\citep{2014MNRAS.445..437M}.

To overcome the redshift-matching problem, \citet{2012PhRvD..85l3510C} have
applied a local regression technique to the SNIa $\dL$ data at redshift
windows of interest with adjustable bandwidth.  However, this method is not
easily generalized to highly correlated data.  It still rejects the majority
of data points outside of the narrow windows, and one might overlook their
influence on $\dL$ estimates through their systematic correlations with data
points inside the windows.

Here, we address the issue by using a Bayesian statistical method detailed in
\citet[][hereafter \citetalias{2016MNRAS.463.1651M}]{2016MNRAS.463.1651M},
which compresses correlated luminosity distance data at given control points
in log-redshift.  This has been specifically developed for the analysis of the
SNIa data from the Joint Light-curve Analysis
\citep[JLA;][]{2014AA...568A..22B}. The goal of this work is to provide
up-to-date, straightforward, and independent measurements of $\eta$ at
selected redshifts, along with their correlations, using $\dL$ from the
compressed JLA data set and $\dA$ estimates from BAO measurements.  The
derived constraints are largely limited by the uncertainties on the
cosmological parameters that the $\dA$ values depend on.  Combining external
information from \textit{Planck} measurements of the cosmic microwave
background (CMB) temperature and polarization anisotropy power spectrum
\citep{2016AA...594A..13P} can significantly reduce such uncertainties.
However, as we will amply explain, \textit{Planck}-derived constraints on
cosmological parameters implicitly assume that there is no violation of the
photon number conservation. In such a case, the DD test can be used as probe
of systematic effects affecting the luminosity distance or the
angular-diameter distance measurements.

In Section~\ref{sec:datasets}, we derive expressions for $\dL$
and $\dA$ in terms of the data.  In Section~\ref{sec:methods}, we describe the
Monte Carlo (MC) analysis methods.  The results are presented and analyzed in
Section~\ref{sec:results} and further discussed in
Section~\ref{sec:discussion}.

\section{Data Sets}
\label{sec:datasets}

\subsection{BAO Angular-diameter Distance}
\label{sec:bao}

We use $\dA$ estimates from BAO measurements of the WiggleZ survey
\citep{2011MNRAS.415.2892B, 2012MNRAS.425..405B} and the Baryon Oscillation
Spectroscopic Survey (BOSS) DR12 \citep{2017MNRAS.470.2617A} consensus
compilation.  WiggleZ data consist of BAO volume distance parameter $A(z)$
\citep{2005ApJ...633..560E, 2011MNRAS.415.2892B} and the Alcock--Paczy{\'n}ski
effect parameter $F(z)$ \citep{1996MNRAS.282..877B, 2011MNRAS.418.1725B} at
effective redshifts 0.44, 0.60, and 0.73, respectively.  From the
measurements, $\dA$ is derived through the following equation
\begin{equation}
    d_{\text{A,WiggleZ}}(z) = \frac{c}{H_0} \frac{A(z) [z^2 F(z)]^{1 /
    3}}{\sqrt{\densm} (1 + z)},
    \label{eq:dawigg}
\end{equation}
where $c$ is the speed of light, $H_0$ is the Hubble constant, and $\densm$ is
the matter density parameter.

BOSS data on the other hand provide consensual estimates of the ratio
$\widetilde{d}_M = \dA (1 + z) \rd^{\text{fid}} / \rd$ by joining the
constraints from BAO features and the full shape of galaxy correlations at
effective redshifts 0.38, 0.51, and 0.61. Here, $\rd$ is the acoustic horizon
at the photon--baryon drag epoch and the BOSS DR12 fiducial value is
$\rd^{\text{fid}} = 147.78$~Mpc. Thus, $\dA$ is expressed in terms of the data
by
\begin{equation}
    d_{\text{A,BOSS}}(z) = \frac{\rd}{\rd^{\text{fid}}}
    \frac{\widetilde{d}_M(z)}{(1 + z)}.
    \label{eq:daboss}
\end{equation}

We notice that the overlap of WiggleZ and BOSS survey volumes makes the two
BAO data sets correlated \citep{2016MNRAS.455.3230B}.  However, a full
analysis consistently joining the data sets is beyond the scope of this work. 

Evidently from \erefs{eq:dawigg} and (\ref{eq:daboss}), in order to use the
BAO data, it is necessary to specify the cosmological parameters, or
``complementary parameters'' (CPs), namely $\vv{\varphi} = (H_0, \densm,
\rd)$.  The CPs are similar to the role of prior distributions in the context
of inference problems, in that they are specified independently of and in
complement to the data to express our belief or uncertainty.  However, unlike
prior distributions, they cannot be updated by the analysis.  In
Section~\ref{sec:methods} we describe the choice of such CPs in detail.

\subsection{SNIa Luminosity Distance}
\label{sec:sn}

We compute compressed SNIa luminosity distance moduli $\mulc$ and their
covariance matrix from the JLA data set with the method detailed in
\citetalias{2016MNRAS.463.1651M}.  The redshifts of compression, or ``control
points,'' are chosen with two criteria in mind.  First, $\mulc$ must be
available at the redshift of $\dA$ data.  Second, the control points should be
distributed such that the statistical uncertainties on $\mulc$ are evenly
imputed to them.  In practice, we perform two separate compression runs for
SNIa + BOSS and SNIa + WiggleZ respectively.  Each compressed data set
contains 15 suitably chosen control points between $0.01 \le z \le 1.30$, and
the relevant data portions are listed in Table~\ref{tab:sn}.  We have verified
that the compression results are not affected significantly by the choice of
other control points.

\begin{deluxetable}{cCcccc}
    \tablecolumns{6}
    \tablecaption{Compressed SNIa Data at Anchoring and BAO-matching Control
    Points \label{tab:sn}}
    \tablehead{\colhead{$z$} & \colhead{$\mulc$} &
    \multicolumn{4}{c}{$\text{Covariance} \times 10^4$}}
    \startdata
    \sidehead{SNIa + BOSS}
    0.01  & 33.12 \pm 0.05 & 21.18   & 2.392   & 1.625   & 1.395 \\
    0.38  & 41.57 \pm 0.03 & \nodata & 8.925   & 0.192   & 2.450 \\
    0.51  & 42.30 \pm 0.03 & \nodata & \nodata & 10.05   & 1.869 \\
    0.61  & 42.74 \pm 0.03 & \nodata & \nodata & \nodata & 12.02 \\
    \sidehead{SNIa + WiggleZ}
    0.01  & 33.12 \pm 0.05 & 21.69   & 2.115   & 1.341   & 0.067 \\
    0.44  & 41.93 \pm 0.03 & \nodata & 8.636   & 1.023   & 3.238 \\
    0.60  & 42.70 \pm 0.03 & \nodata & \nodata & 9.746   & 4.636 \\
    0.73  & 43.21 \pm 0.05 & \nodata & \nodata & \nodata & 26.00
    \enddata
    \tablecomments{Covariance values have been scaled by $10^4$ for
    presentation.}
\end{deluxetable}

It should be noted that the compression step computes the distance moduli only
up to an implicit magnitude offset $M$.  It is the quantity
\begin{equation}
    \mulc = \muL - M =  5 \logg \left( \frac{\dL}{10~\text{pc}} \right) - M
    \label{eq:defdistmod}
\end{equation}
that is produced by the compression.  The parameter $M$ in
\eref{eq:defdistmod} is degenerate with $H_0$ as discussed in
\citetalias{2016MNRAS.463.1651M} \citep[see also][]{2013ApJ...777L..24Y}, and
it is possible to eliminate $M$ and to obtain $\dL$ that is directly
comparable to $\dA$.  We exploit the fact that at the lowest available, or the
``anchoring'' redshift $z_1 = 0.01$, the luminosity distance can be
approximated to the second order as
\begin{equation}
    \dL(z_1) = \frac{c z_1}{H_0} \left[ 1 + \frac{1}{2} \left( 1 - q_0
    \right) z_1 + \mathcal{O}(z_1^2) \right] \approx \frac{c z_1}{H_0},
    \label{eq:dlapprox}
\end{equation}
where $q_0$ is the deceleration parameter.  For the small value of $z_1$,
higher-order terms in \eref{eq:dlapprox} are negligible (unless one must
consider unrealistic cosmological scenarios with $\left\vert q_0 \right\vert
\approx 10^2$).  This allows us to express $M$ by $\dL(z_1) \approx c z_1 /
H_0$ and $\mulc(z_1)$ using \eref{eq:defdistmod}.  Carrying out the algebra,
we obtain the expression for $\dL(z)$ as
\begin{equation}
    \dL(z) = \left( \frac{c z_1}{H_0} \right) 10^{\left[ \mulc(z) -
    \mulc(z_1) \right] / 5}.
    \label{eq:dlfixed}
\end{equation}

\section{Methods}
\label{sec:methods}

We derive the probability density function (PDF) of $\eta$ at a given redshift
from MC samples of $\dL$ and $\dA$ inferred from the observational data sets.
The underlying idea is that the SNIa and BAO distance data, the CPs, and
$\eta$ at the chosen redshifts are all random variables.  In particular,
$\eta$ is a transformation of the combined random variable of data and CPs,
which is specified by the composition of \eref{eq:dd} with \eref{eq:dlfixed}
and either \eref{eq:dawigg} or (\ref{eq:daboss}) for WiggleZ or BOSS data,
respectively.

The view of observational data as random variables fits naturally into the
Bayesian statistical inference framework commonly encountered in the study of
cosmological models \citepalias[for example,
see][Section~2]{2016MNRAS.463.1651M}.  The CPs themselves are often obtained
from Bayesian inference with observational data.  In such a case, a
self-consistent analysis demands that the inference of CPs does not rely on
the $\dA$ and $\dL$ data sets used here, and that the underlying statistical
model used in the CP inference does not put restrictive assumptions on $\eta$
or related functions.

In practice, it can be difficult to unambiguously satisfy both these points,
and we must also be attentive to the context of their validity.  Still, we can
make our best efforts in this direction. 

\subsection{Complementary Parameters}
\label{sec:cps}

Following the discussion in Section~\ref{sec:bao}, in order to estimate the
BAO angular-diameter distances, we need input on the CPs, $\vv{\varphi} =
(H_0, \densm, \rd)$, while for SNIa data we need to incorporate the dependence
on $H_0$.  In this work, we consider two CP sets motivated by current
knowledge of those parameters from independent observations.

The first CP set consists of a joint distribution on $(h, \densm, \densb h^2)$
where $h = H_0 / 100~\text{km~s}^{-1}~\text{Mpc}^{-1}$ is the dimensionless
Hubble constant and $\densb h^2$ the baryon energy density parameter.  We
sample $h$ from a conservative choice, namely the Gaussian distribution
$\mathcal{N}(0.688, 0.033^2)$ used in \citetalias{2016MNRAS.463.1651M} based
on the re-selected and re-calibrated nearby SNIa distances with the
independent megamaser distance to NGC 4258 as the Cepheid zero point
\citep{2015ApJ...802...20R}.  The usefulness of NGC 4258 as a calibration
source with independent, well-understood systematic uncertainties is explained
by \citet{2014MNRAS.440.1138E}.  We further adopt the conservative estimate
$\densb h^2 \sim \mathcal{N}(0.02228, 0.00084^2)$ from a big-bang
nucleosynthesis analysis with relic \nuclide[4][]{He} and deuterium abundance
data \citep[][table~V]{2016RvMP...88a5004C}. For $\densm$, we use a simple,
non-informative distribution, namely the uniform distribution over the range
[0.15, 0.45], which is inclusive enough to cover independent constraints from
the mass function of galaxy clusters \citep{2015ApJ...799..214B}.
Furthermore, as an indirect check on these choices, we compute the baryon
fraction $\fb = \densb / \densm$ as implied by the random samples.  The
resultant $\fb$ distribution, with mean and standard deviation 0.17 $\pm$
0.06, is consistent with independent constraints from galaxy cluster
observations \citep{2013ApJ...778...14G, 2016MNRAS.455..258C}.

In order to check the sensitivity of the results to the particular choice of
$h$, we also derive constraints by sampling $h$ from a Gaussian distribution
with $\mathcal{N}(0.7348, 0.0166^2)$ based on the Cepheid period--luminosity
relation determined by parallaxes of Milky Way Cepheids
\citep{2018ApJ...855..136R}.  The other parameters' distributions are
unmodified.  We will refer to this alternative CP set as ``H73.''

From these random samples, we derive the sample for $\rd$, which is necessary
for application with BOSS BAO distances.  Following the discussions in
\citet{2012MNRAS.427.2168M} and \citet{2014MNRAS.441...24A}, we evaluate $\rd$
as a function of $(h, \densm, \densb h^2)$ using the software CAMB
\citep{2000ApJ...538..473L, cambzenodo} with the other cosmological parameters
fixed at the values of the BOSS fiducial \lcdm model specified in
\citet{2017MNRAS.470.2617A}. The CP set thus generated is denoted by the label
``Synthetic'' in the rest of this paper, for it is based on the combination of
independent observation constraints.  The sample size is $2 \times 10^6$.

The other CP choice is based on the Markov chain MC analysis for the
Bayesian cosmological parameter constraints of the KiDS-450 tomographic weak
lensing (WL) survey \citep{2017MNRAS.465.1454H}, including posterior
samples\footnote{\url{http://kids.strw.leidenuniv.nl/sciencedata.php}} for
$H_0$, $\densm$, and $\rd$.  They are valuable as an independent,
data-informed source for $\densm$ and $\rd$.  However, WL alone offers no
informative update on its $H_0$ prior.  If one had accepted the $H_0$
constraint as it is, value ranges far removed from informative observational
measurements \citep[such as][]{2016ApJ...826...56R, 2018ApJ...855..136R,
2017Natur.551...85A} would have been over-weighted. For this reason, we
perform a re-weighting of the Markov chains by a weighting function $f_h$, the
Gaussian PDF underlying the $h$ distribution in the Synthetic CP set.  The
re-weighting is implemented with an accept--reject MC algorithm.  For each
sample point in the KiDS-450 Markov chain output, it is randomly accepted with
the suitably normalized probability $p \propto f_h$.  Overall, the acceptance
rate is about 32.6\%, leaving a sample size of about $6.9 \times 10^5$.  We
have verified that the induced shifts in the distributions of $\densm$ and
$\rd$ are about $0.1 \sigma$.  This confirms that the $h$-based re-weighting
does not contaminate the relevant WL-inferred cosmological parameters
noticeably.  We thus obtain an alternative CP set, and for simplicity, in the
following sections we refer to it as ``KiDS.''

\subsection{Cosmic Distance Samples}
\label{sec:obsdist}

We now discuss the random samples of $\dL$ and $\dA$ generated from the
observational data sets described in Section~\ref{sec:datasets}.  Their
distributions are well-approximated by multivariate Gaussian random variables.
We generate the two samples separately and verify that they are not correlated
with the CP samples.  In the case of the WiggleZ BAO data, we generate the
$(A, F)$ joint Gaussian sample using the mean vector and covariance matrix of
\citet{2012MNRAS.425..405B}, having marginalized over the growth rate
parameter $f \sigma_8$. This is combined with the CP samples through
\eref{eq:dawigg} to obtain the sample of $d_{\text{A,WiggleZ}}(z)$.  In the
case of BOSS data, the Gaussian sample of $\widetilde{d}_M$ is created using
the mean and covariance
values\footnote{\url{https://data.sdss.org/sas/dr12/boss/papers/clustering/}}
of \citet{2017MNRAS.470.2617A}, having marginalized over the $\rd$-scaled
expansion rate $(\rd / \rd^{\text{fid}}) H$ and $f \sigma_8$.  Then, by
combining through \eref{eq:daboss} the $\widetilde{d}_M$ sample with that of
$\rd$, we obtain the sample of $d_{\text{A,BOSS}}(z)$.

For the SNIa data, we generate Gaussian samples of $\mulc$ based on the
compressed distance moduli described in Section~\ref{sec:sn}.  Combining them
with $H_0$ through \eref{eq:dlfixed}, we obtain the $\dL$ samples.

Finally, by combining the $\dA$ and $\dL$ samples, we derive $\eta$ through
its definition in \eref{eq:dd} as two distinct samples from SNIa + WiggleZ and
SNIa + BOSS respectively. In each case, the data sample size is matched with
the CP sample.  It is worth noticing that the distance scale $c / H_0$ is
eliminated by combining \erefs{eq:dawigg} and (\ref{eq:dlfixed}).  As a
result, the $\eta$ distribution from SNIa + WiggleZ is independent of $H_0$.
This is not true for SNIa + BOSS, because $\rd$ deviates from the scaling $\rd
\propto H_0^{-1}$ due to the effect of cosmic expansion rate on early-Universe
matter-to-radiation ratio \citep{1995PhRvD..52.5498H} and recombination rates
\citep{2000ApJS..128..407S}.

\section{Results}
\label{sec:results}

\subsection{Testing the DD Relation}
\label{sec:res-ddtest}

\begin{figure}
    \plotone{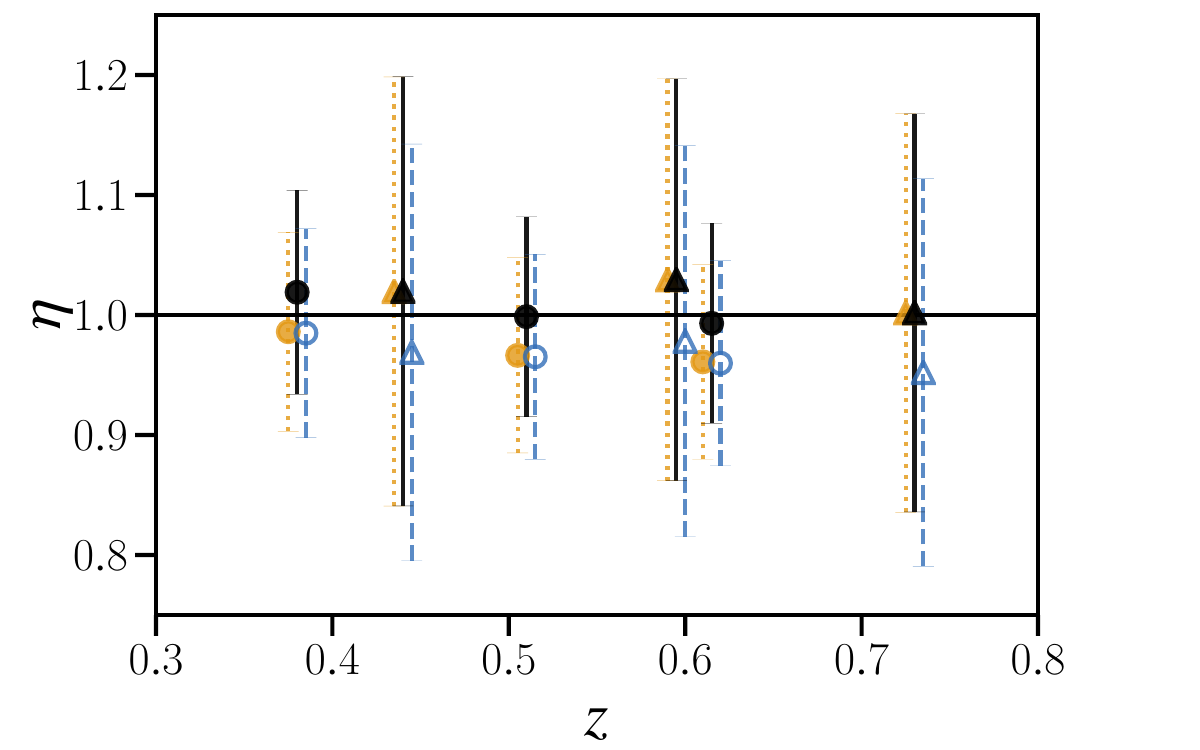}
    \caption{Marginal mean and standard deviation of $\eta$.  Results with the
	Synthetic CP are shown in black as filled markers with solid error
	bars, and those with the KiDS CP are shown in blue and as open markers
	with dashed error bars.  Results with H73 are shown in orange
	(dotted). SNIa + BOSS results are shown as circles, and those
	from SNIa + WiggleZ as triangles.  For readability, the redshift
	locations are shifted slightly around their actual values.
    \label{fig:margeta}} 
\end{figure}

We derive constraints on $\eta$ from the analysis of SNIa + BOSS and SNIa +
WiggleZ samples separately.  Figure~\ref{fig:margeta} shows the mean and
standard deviation estimated from the random samples, and the corresponding
values are quoted in Table~\ref{tab:respm}.  In Figure~\ref{fig:margeta}, we
also show the constraints inferred using H73. As we can see, despite the
discrepancy between the choices of $h$ distribution, it has no significant
effect on the inferred bounds on $\eta$. Here, we stress again that results
from different BAO surveys cannot be combined trivially.  In the case of the
KiDS CP, because the analysis partially depends on Markov chains, we have used
the method of batch means \citep{f08} to verify that the $\eta$ samples
provides sufficiently accurate sample statistics and that the values of the
mean and standard deviation reported here do not exceed their significant
figures.

\begin{deluxetable}{cCCCC}
    %\tablewidth{0.75\columnwidth}
    \tablecaption{Statistical Summary of the DD Test Random Variable
    $\eta$\label{tab:respm}}
    \tablehead{\colhead{$z$} & \colhead{Synthetic\tablenotemark{a}} &
    \colhead{KiDS\tablenotemark{a}} & \colhead{Synthetic\tablenotemark{b}} &
    \colhead{KiDS\tablenotemark{b}}}
    \def\arraystretch{0.8}
    \startdata
    \sidehead{SNIa + BOSS}
    0.38  & 1.02 \pm 0.08 & 0.98 \pm 0.09 &
    1.06^{+0.06}_{-0.13} & 0.96^{+0.10}_{-0.08} \\
    0.51  & 1.00 \pm 0.08 & 0.96 \pm 0.08 &
    1.04^{+0.06}_{-0.12} & 0.95^{+0.10}_{-0.08} \\
    0.61  & 0.99 \pm 0.08 & 0.96 \pm 0.08 &
    1.03^{+0.06}_{-0.12} & 0.94^{+0.10}_{-0.08} \\
    \sidehead{SNIa + WiggleZ}
    0.44  & 1.02 \pm 0.18 & 0.97 \pm 0.17 &
    1.06^{+0.14}_{-0.24} & 0.92^{+0.20}_{-0.16} \\
    0.60  & 1.03 \pm 0.17 & 0.98 \pm 0.16 &
    1.12^{+0.10}_{-0.27} & 0.93^{+0.19}_{-0.15} \\
    0.73  & 1.00 \pm 0.17 & 0.95 \pm 0.16 &
    1.08^{+0.10}_{-0.26} & 0.91^{+0.19}_{-0.15} 
    \enddata
    \def\arraystretch{1.0}
    \tablenotetext{a}{Mean and standard deviation.}
    \tablenotetext{b}{Mode and 68.3\% credible interval.}
\end{deluxetable}

We find the $\eta$ sample distributions to be skewed. Hence, the statistical
uncertainties can be characterized more precisely by a mode and credible
interval analysis. To this end, we use a Gaussian kernel density estimator to
overcome MC noise and smooth the sample distribution.  Then, we find the
approximate location of the mode for the smoothed one-dimensional marginal
distribution at each redshift. The mode estimates and the $68.3\%$ credible
intervals\footnote{We compute the approximate credible interval $[a, b]$ for
given probability level $p$ such that $f_G(a) = f_G(b)$ and $\int_a^b f_G(x)
dx = p$, where $f_G$ is the smoothed sample PDF. The credible interval thus
defined intuitively follows the concept of the Lebesgue integral and is useful
for describing the asymmetric shape. Moreover, as can be proved using a
Lagrange multiplier, it is a minimal one for unimodal $f_G$ with strictly
monotonous wings separated by the mode.} are quoted in Table \ref{tab:respm}.

\begin{figure*}
    \gridline{
	\fig{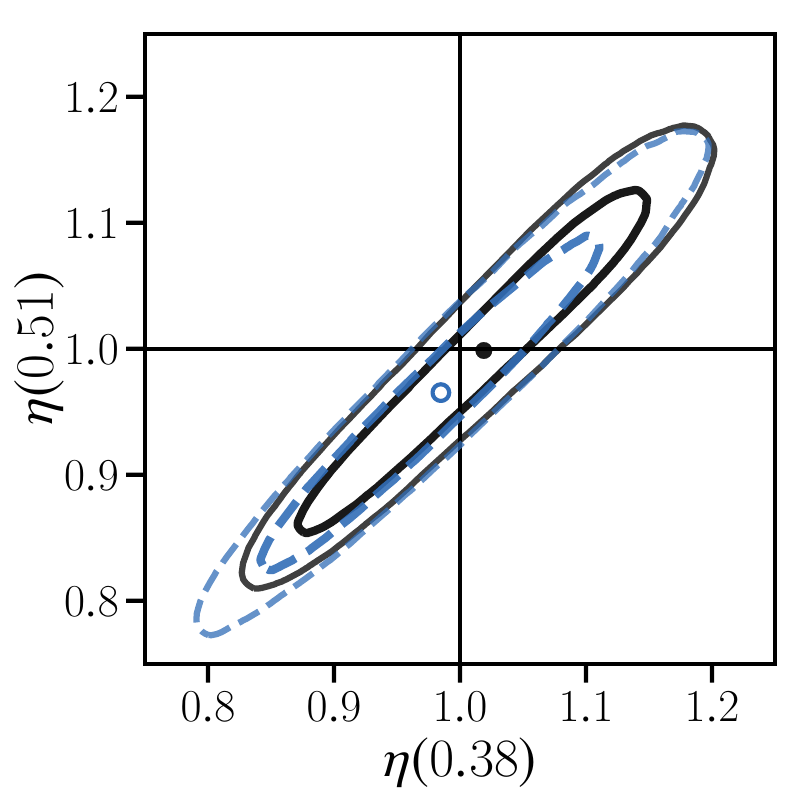}{0.32\textwidth}{\label{fig:bosseta_a}}
	\fig{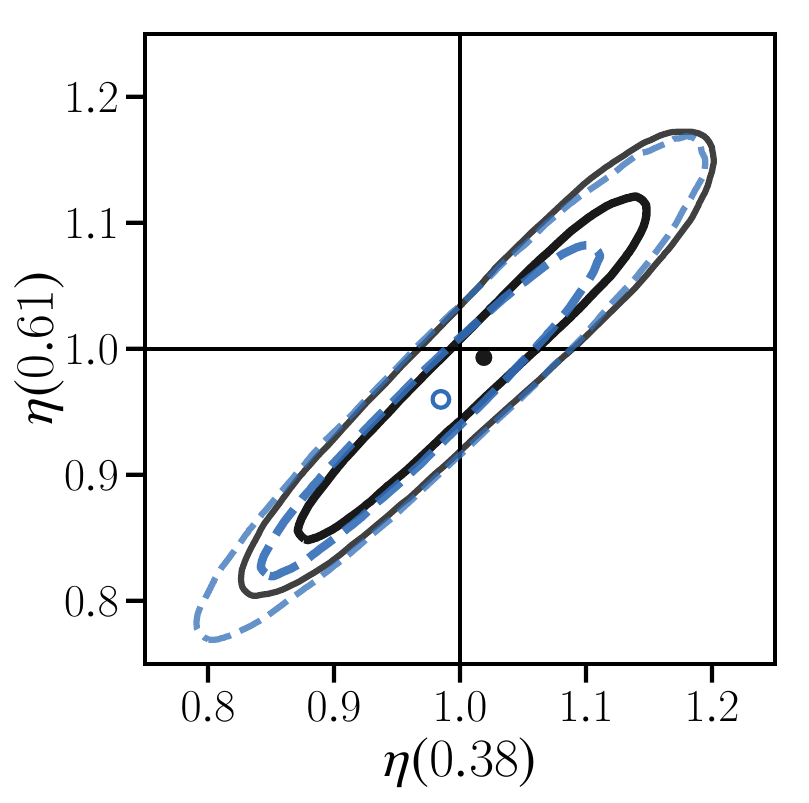}{0.32\textwidth}{\label{fig:bosseta_b}}
	\fig{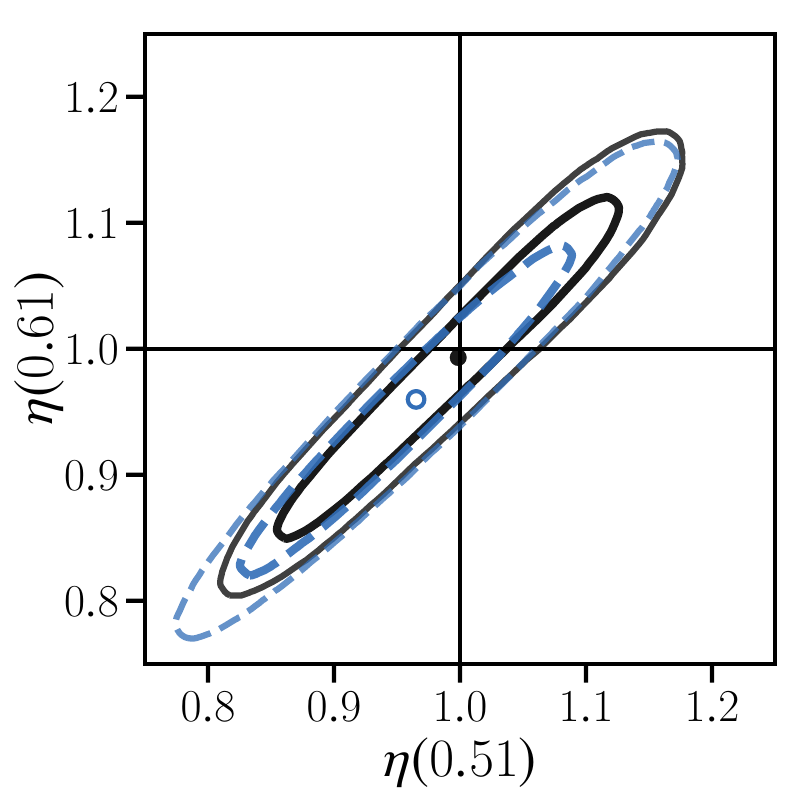}{0.32\textwidth}{\label{fig:bosseta_c}}
    }
    \caption{Two-dimensional MC distribution of $\eta$ from SNIa + BOSS.  Each
	contour set encloses a region measuring $p_1 = 0.683$ (inner, thick)
	and $p_2 = 0.954$ (outer, thin) under the corresponding sample
	distribution.  Solid (black) contours show the results with Synthetic
	CPs, and the dashed (blue) ones with the KiDS.  The mean value for
	each case is shown by a marker of the matching
	color.\label{fig:bosseta}}
\end{figure*}

\begin{figure*}
    \gridline{
	\fig{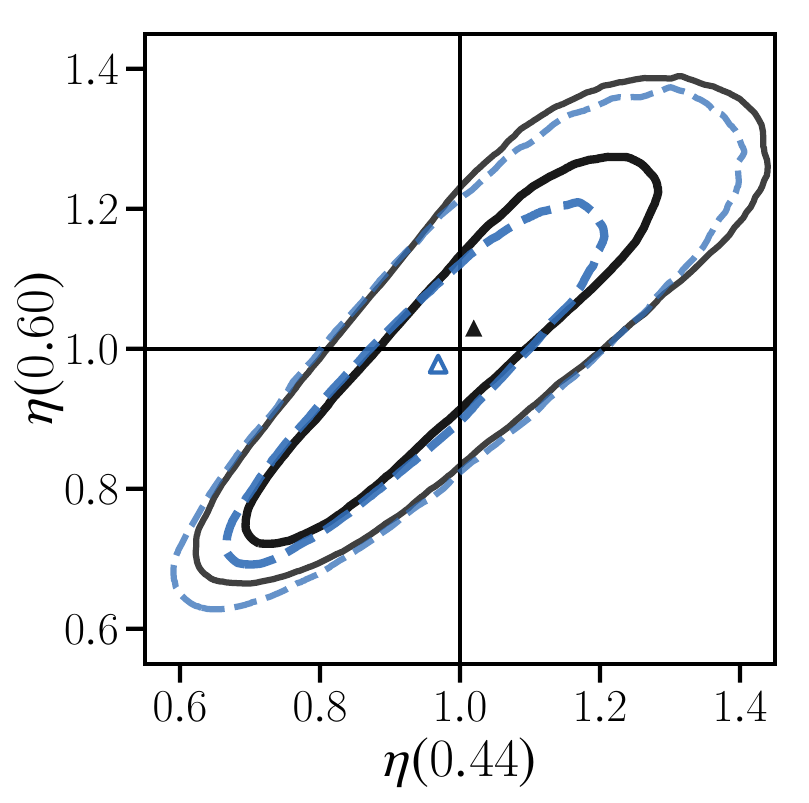}{0.32\textwidth}{\label{fig:wiggeta_a}}
	\fig{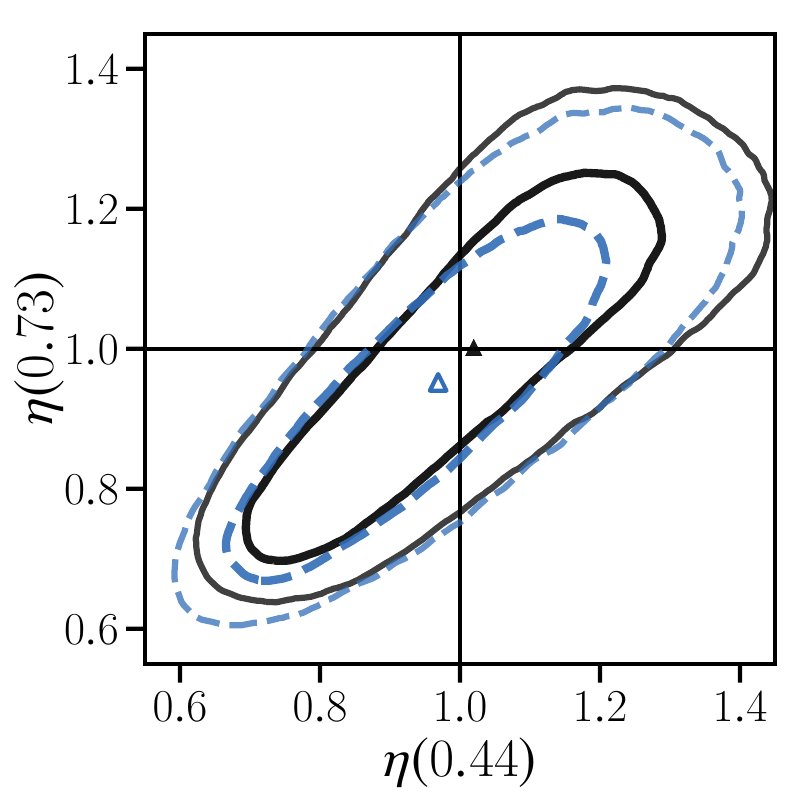}{0.32\textwidth}{\label{fig:wiggeta_b}}
	\fig{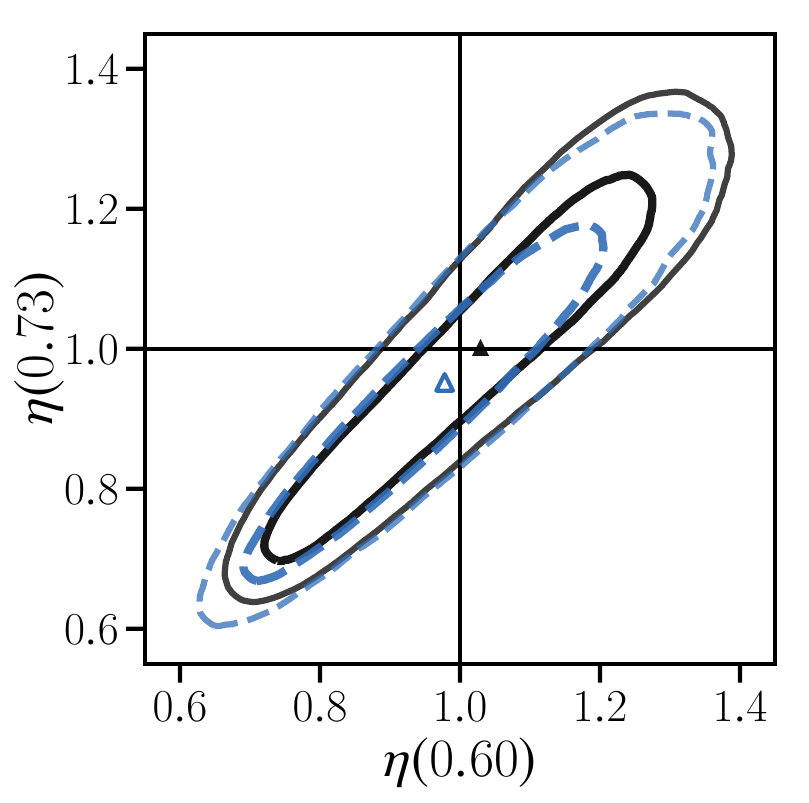}{0.32\textwidth}{\label{fig:wiggeta_c}}
    }
    \caption{Similar to Figure~\ref{fig:bosseta} but from SNIa + WiggleZ and
	with different scales.\label{fig:wiggeta}}
\end{figure*}

The $\eta$ distributions obtained from this analysis are correlated from one
redshift to another. This can be better appreciated in
Figures~\ref{fig:bosseta} and \ref{fig:wiggeta}, which show the
two-dimensional joint constraints from SNIa + BOSS and SNIa + WiggleZ,
respectively.

These results indicate the absence of substantial evidence for deviations from
\eref{eq:dd}. Moreover, there is no clear trend of $\eta(z)$ evolution.

From Section~\ref{sec:methods}, we can readily understand that the large
statistical uncertainties on $\eta$ are a consequence of the quality of both
the distance data and the CPs. Tighter CPs can be used at the cost of
generality, and as previously noted, when testing the DD relation, one must
pay attention to the assumptions under which the CPs have been derived. In
particular, one may be inclined to include one of the most stringent
constraints on the cosmological parameters, namely the results from
\textit{Planck} measurements of CMB anisotropy power spectra
\citep{2016AA...594A..13P}. However, such results implicitly assume the photon
number conservation and the validity of the DD relation. Any process violating
the photon number conservation during the photon--baryon coupling epoch or the
propagation of CMB photons will likely induce temperature anisotropy and
modify the power spectra, eventually leading to a different cosmological
parameter inference.  Unfortunately, the effects of photon-number violating
processes on CMB are highly model-specific \citep[see,
e.g.,][]{2016JCAP...04..050R}.  As such, tight constraints on $\eta$ obtained
by including CMB information are difficult to interpret \citep[see
also][]{2014MNRAS.443.1881C}.

However, this does not imply that the incorporation of CMB constraints (and
implicitly their assumptions) cannot lead to a meaningful comparison between
the SNIa and BAO cosmological distances. In fact, one can assume the DD
relation to be valid and use the inferred constraints on $\eta$ as a proxy of
potential systematics affecting cosmic distance estimations.  The nature of
these systematic effects does not have to be exotic physics to which CMB
anisotropies are sensitive, but rather the result of unaccounted for yet
mundane mechanisms independent of the CMB.  As an example, in
\citet{2016PDU....14...57E} the validity of DD was used to test the
calibration of the SNIa standard-candle relation.  Hereafter, we will present
results on cosmic distance systematics using the DD estimates in combination
with a CMB-informed CP.

\subsection{Cosmic Distance Systematics}
\label{sec:ddsystematics}

In the following, we assume the DD relation to hold and use the estimates of
$\eta$ from SNIa and BAO in combination with \textit{Planck} results to derive
constraints on systematics affecting cosmic distance measurements. In
particular, we take the CPs $(H_0, \densm, \rd)$ (see Section~\ref{sec:cps}),
from the posterior Markov chains of the flat \lcdm ``base'' model parameters
obtained from the \textit{Planck} TT + TE + EE + low-$\ell$ temperature and
polarization (``lowP'') anisotropy.\footnote{The chain files were downloaded
from the \textit{Planck} Legacy Archive (\url{http://pla.esac.esa.int/pla/}).
The ones used here are from the directory
\texttt{base\_plikHM\_TTTEEE\_lowTEB}.} Again, we have checked that there is
minimal correlation between the chains and the data samples.  We dub this CP
set as ``\textit{Planck}.'' Its sample size is about $1.07 \times 10^5$.

In the case of SNIa $\dL$ measurements, systematic effects may arise from
a variety of sources \citep[see, e.g.,][]{2011ARNPS..61..251G}. As suggested
in \citet{2006MNRAS.372..191C}, one way of using the estimates on the
deviations from the DD relation is to test the presence of dust extinction due
to an intergalactic gray dust component that is not removed through standard
color analysis. This extinction would systematically dim SNIa, thus making
them appear more remote, but would not affect the BAO distance as indicated by
the shape and location of the acoustic peak in the galaxy correlation
functions.

In such a case, the rest-frame \textit{B}-band extinction correction to the
SNIa standard-candle magnitude, $A_B$, is related to $\eta$ by $A_B(z) -
A_B(z_1) = 5 \logg \eta(z)$, where, again, $z_1 = 0.01$ is the anchoring
redshift (see Section~\ref{sec:sn}).  At that low redshift, the optical depth
and intergalactic extinction is typically negligible.  Therefore, in the
remainder of this paper, we will simply refer to $A_B(z)$.

\begin{figure}
    \plotone{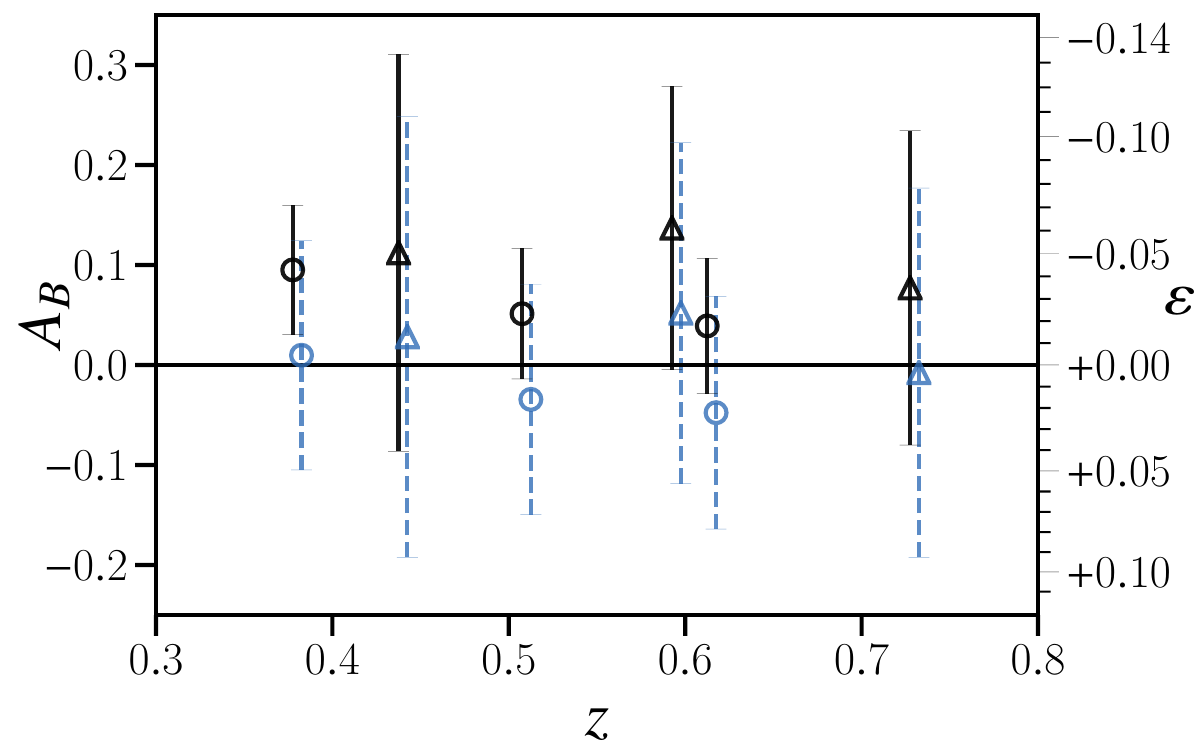}
    \caption{Marginal mean and standard deviation of $A_B$ as a
	function of redshift $z$ assuming $\dA$ values derived from
	\textit{Planck} CPs. Results from the \textit{Planck} and
	\textit{Planck}-$w$CDM CPs are represented, respectively, by black
	(solid) and blue (dashed) bars.  Similar to Figure~\ref{fig:margeta},
	circles and triangles denote the result from SNIa + BOSS and SNIa +
	WiggleZ, respectively.  On the right side, an alternative set of
	scales shows the same results expressed in terms of $\varepsilon$ as a
	proxy of systematics affecting BAO angular-diameter distances. The
	spacings on the $\varepsilon$ scale are logarithmic in $(1 +
	\varepsilon)$, but within the relevant data range, they are visually
	indistinguishable from linear scales.
    \label{fig:abextinction}}
\end{figure}

Table~\ref{tab:ab} displays the mean and standard deviation of the extinction
$A_B$. These values are shown at their respective redshifts in
Figure~\ref{fig:abextinction}. Again, batch means are used to check the
accuracy of these results. The $A_B$ samples are sufficiently symmetric when
marginalized to each redshift, and the credible interval analysis reveals no
substantial difference from $1\sigma$ bounds.  As a joint distribution, the
SNIa + BOSS result is closely approximated by the multivariate Gaussian.  For
future reference, we report the tests for normality and the MC estimates for
its mean vector and covariance matrix in Appendix~\ref{sec:a:ab}.

\begin{deluxetable}{cCC}
    \tablecaption{Mean and Standard Deviation of the Extinction Correction
    $A_B$ Marginalized at Each Redshift Using \textit{Planck}-based CP
    Sets. \label{tab:ab}}
    \tablehead{
	\colhead{$z$} & \colhead{\textit{Planck}} &
	\colhead{\textit{Planck}-$w$CDM}}
    \startdata
    \sidehead{SNIa + BOSS}
    0.38  & 0.10 \pm 0.06 & \phm{-}0.01 \pm 0.11 \\
    0.51  & 0.05 \pm 0.06 &       -0.03 \pm 0.12 \\
    0.61  & 0.04 \pm 0.07 &       -0.05 \pm 0.12 \\
    \sidehead{SNIa + WiggleZ}
    0.44  & 0.11 \pm 0.20 & \phm{-}0.03 \pm 0.22 \\
    0.60  & 0.14 \pm 0.14 & \phm{-}0.05 \pm 0.17 \\
    0.73  & 0.08 \pm 0.16 &       -0.01 \pm 0.18 
    \enddata
\end{deluxetable}

As we can see, the overall results are consistent with a null extinction
magnitude, although there appears to be a slight preference of the sign, $A_B
\ge 0$ (see Appendix~\ref{sec:a:ab} for a more detailed explanation).

The results described above constitute the main results for $A_B$
based on the \textit{Planck} base \lcdm model that is precisely constrained by
the CMB power spectra.  To check whether an alternative dark energy
prescription might modify our interpretation, we perform a similar analysis
based on the \textit{Planck}-$w$CDM cosmological posterior constraints.  

It is worth noticing that a general problem with CMB constraints and non-\lcdm
dark energy is parameter degeneracy.  As the CMB spectral features are only
indirectly sensitive to late-time evolution, free
parameters introduced to describe more complex dark energy may not be
well-constrained, and degeneracies may arise among parameters \citep[see
also][]{2016AA...594A..14P}.  In the case of $w$CDM, the Hubble constant $h$
fails to be constrained, as it is the case with KiDS posterior analysis.
Therefore, we adopt the same re-weighting by the conservative Gaussian
distribution (see Section~\ref{sec:cps}).  The inability to constrain $h$ has
also be addressed by the \citet[][Section~5.4]{2016AA...594A..13P}, and our
re-weighting distribution is an update from their ``conservative prior'' based
on \citet{2014MNRAS.440.1138E}.

\begin{figure}
    \plotone{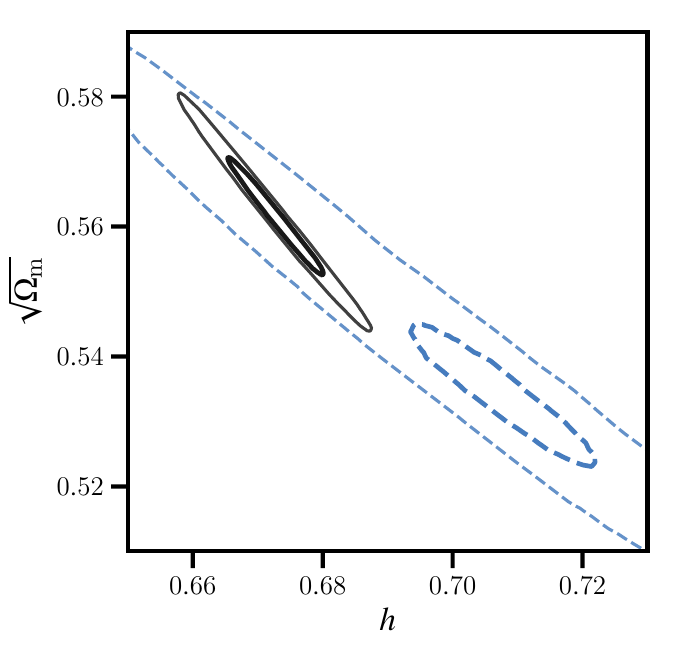}
    \caption{Shift of parameters $(h, \sqrt{\densm})$ from \textit{Planck}
	base \lcdm (solid, black) to $w$CDM (dashed, blue) posteriors, shown
	as $p_1 = 0.683$ and $p_2 = 0.954$ credible regions.
    \label{fig:wdegen}}
\end{figure}

The results are displayed in tandem with the \textit{Planck} base results in
Table~\ref{tab:ab} and Figure~\ref{fig:abextinction}. The use of re-weighted
\textit{Planck}-$w$CDM CP shifts $A_B$ closer to zero, but the standard
deviation is about twice as large as the \lcdm one in the case of SNIa + BOSS
even after re-weighting. The smaller sample size ($1.0 \times 10^4$) also
causes larger MC standard errors, but they remain dominated by the
distributional spread.  The shift results from the fact that $\rd$ (a
parameter not directly affected by late-time dark energy) remains almost
unchanged from the base \lcdm distribution, while the combination of $(h,
\sqrt{\densm})$ shifts along the direction of parameter degeneracy, shown in
Figure~\ref{fig:wdegen}.  Overall, any evidence of deviation from $A_B = 0$ is
further weakened.

To compare with the earlier works, we cast the results in terms of the optical
depth
\begin{equation}
    \tau = \frac{(\ln 10)}{2.5} A_B = 2 \ln \eta
    \label{eq:odepth}
\end{equation}
and take its redshift differential, thereby eliminating all dependence on the
CPs.  Using the SNIa + BOSS data set, we obtain $\tau(0.51) - \tau(0.38) =
-0.04 \pm 0.05$ and $\tau(0.61) - \tau(0.38) = -0.05 \pm 0.05$.  We have
verified that these differentials, as expected, are essentially the same up to
a small sampling error, independent of the CP choice.  The uncertainties on
$\Delta \tau$ are lower than in previous studies \citep{2009ApJ...696.1727M,
2012JCAP...12..028N} by virtue of higher-precision distance data.  Meanwhile,
there is no conclusive support for evolving $\tau(z)$.

The generality of estimating relative increments in $\tau$ is gained at the
cost of losing information about its amount in absolute terms.  In contrast,
our main $A_B$ estimates can be compared with independent estimations
of intergalactic extinction.  At $z = 0.38$, we find our result broadly
consistent with $A_B \approx 0.02$ reported by \citet{2010MNRAS.406.1815M,
2010MNRAS.405.1025M} from $z \approx 0.36$ and other observational constraints
cited therein.

These bounds are only one possible interpretation of cosmic distance
systematics that induce apparent deviations from the DD relation.  Should one
uses the SNIa $\dL$ to access possible systematic shifts in BAO $\dA$, one
would have used $\varepsilon = \eta^{-1} - 1$ to express the increment by
which the BAO-measured $\dA$ shifts relative to $\dL / (1 + z)^2$, which we
show as the right-side scale in Figure~\ref{fig:abextinction}.

\section{Discussion}
\label{sec:discussion}

In this work, we have performed DD relation tests using recent SNIa and BAO
data.  Assuming auxiliary cosmological information (i.e., CPs) that is
necessary to obtain comparable $\dA$ and $\dL$ values, we find a 9\%
constraint consistent with $\eta = 1$ from the analysis of SNIa + BOSS.  The
combination SNIa + WiggleZ is affected by greater statistical uncertainties in
the BAO distances, but it allows us to probe a different redshift range, and
we obtain qualitatively similar results with about 18\% uncertainty in $\eta$.

Our results stand in contrast to earlier analyses using SNIa + clusters
\citep[e.g.,][]{2004PhRvD..70h3533U, 2010ApJ...722L.233H} or SNIa + BAO
\citep[e.g.,][]{2012JCAP...12..028N}, in which $\eta < 1$, or anomalous
brightening, was reported as being close to or above $1\sigma$ level.  We
suspect the origin of their seemingly surprising conclusion might partially
lie in the difference in the methods of statistical analysis.

The inclusion of tighter-bounded CPs, such as those from the \textit{Planck}
\lcdm CMB analysis, would lead to much tighter constraints on $\eta$
with about $3\%$ uncertainty. However, the \textit{Planck} analysis assumes
photon number conservation. Thus, \textit{Planck} CPs cannot be used to test
the DD relation directly. Nevertheless, such CPs can be combined with SNIa and
BAO data to constrain systematic effects on cosmic distance measurements that
manifest as an apparent deviation from the DD relation.  In this work, we
present examples of such analysis by inferring bounds on the SNIa
extinction.We demonstrate the fact that such analysis dependents on
high-precision cosmological posterior, while parameter degeneracy encountered
with more complex dark energy models should be mitigated.  In future studies,
it will be worth exploring how the issue for precision may be approached in
each context, especially in the presence of difficulty with combining
cosmological information from independent probes in the context of extended
dark energy models \citep[see, e.g.,][]{2016MNRAS.463.1416G}.

The work presented here differs from previous analyses not only by the use of
updated data but primarily by featuring new analysis methods.

The SNIa compression procedure \citepalias{2016MNRAS.463.1651M} produces
accurate data covariance by properly treating the SNIa standardization
uncertainties, in contrast to $\chi^2$ expressions found in
similar studies \citep[e.g.,][]{2015PhRvD..92l3539L} that would be inadequate
for this task.  Meanwhile, the method obviates the need to use narrow bands
for redshift-matching.  Compared with earlier approaches
\citep[e.g.,][]{2009ApJ...696.1727M, 2012PhRvD..85l3510C, 2016JCAP...07..026R},
our compression is done in $\logg z$ space where the systematic evolution of
$\muL$ varies less nonlinearly, allowing us to use larger bandwidths.  This
reduces statistical uncertainties due to limited local sample size and is more
robust against the systematics induced by a possibly nonrepresentative local
sample.  A similar method \citep{2013MNRAS.436.1017L} was used with earlier
Union2 SNIa data \citep{2010ApJ...716..712A}, but it did not share the
aforementioned benefits and was not generalized to non-diagonal data
covariance.  Recently, a smoothing interpolation method based on Gaussian
processes was applied to the analysis of DD relation
\citep{2017JCAP...07..010R}, which could be used to reconstruct $\eta(z)$ as a
smooth function.  However, unlike in ours, in the aforementioned study (unlike
ours), contribution to the final statistical distributions from SNIa
standardization uncertainty was not accounted for.

Another advantage of our approach concerns the estimation of uncertainties on
$\eta$.  MC sampling allows us to directly propagate the probabilistic
uncertainties of the data and the CP onto the distribution of $\eta$ or its
functions consistently, without the need of assuming Gaussian
uncertainties.  In fact, not all of our results can be robustly approximated
as Gaussian (see Appendix~\ref{sec:a:ab}).  Our method can faithfully model
the correlated uncertainties of $\eta$ estimations at different redshifts,
which must be taken into account when investigating possible evolution of
$\eta(z)$ or related quantities (see Section~\ref{sec:ddsystematics}).  To the
authors' knowledge, this is the first time the issue of cross-redshift
correlation is explicitly demonstrated in similar studies.

Finally, unlike previous works, our test does not rely on parametric
constraints of artificial $\eta(z)$ evolution templates.  We indeed find that
no such evolution could be convincingly indicated by current data.  From the
standpoint of statistical methodology, currently the availability of
high-quality, independent, and matching $\dL$ and $\dA$ measurements is still
scarce, thus not allowing many degrees of freedom for parametric fitting. Our
attention thus focuses on the distributional properties of $\eta$ itself. We
leave a parametric characterization of $\eta(z)$ to the future availability of
abundant data. 

In the future, surveys such as
\textit{Euclid}\footnote{\url{http://sci.esa.int/euclid/}} and
\textit{LSST}\footnote{\url{https://www.lsst.org/}} will increase the data
sample size and improve the study of systematic effects in cosmic distance
measurements, thereby allowing us to verify the DD relation with higher
precision and accuracy.  As an approximate evaluation, we assume the
statistical uncertainty on $\mulc$ scales as the inverse square root of SNIa
sample size $N$, and that the uncertainties on BAO and WL $\dA$ at $z \approx
0.5$ remains at the current 1.4\% level, a conservative estimate based on
\textit{Euclid} science objectives
\citep[Section~3.1.1.2]{EuclidScird}.\footnote{\url{http://sci.esa.int/euclid/42822-scird-for-euclid/}}
Assuming further the current \textit{Planck} \lcdm CPs and \textit{LSST} SNIa
``deep'' sample size of $N \approx 10^4$
\citep[Section~11.2.2]{2009arXiv0912.0201L}, we estimate the forecast
statistical uncertainty on $\eta$ at $z \approx 0.5$ to be about 0.02, and
about 0.04 mag on $A_B$, using mock data.  It is worth pointing out that our
method based on local compression of SNIa could be adapted to future
large-sample SNIa surveys, while alternative methods based on individual SNIa
selection or narrow-windowed local regression might exacerbate the effect of
nonrepresentative subsamples (see Section~\ref{sec:intro}). The high precision
of future data may us provide us with more stringent validations of the DD
relation or greater insight into the physical origin of any apparent violation
thereof.

The data files and data-analysis programs used in this work are publicly
available.\footnote{\url{https://doi.org/10.5281/zenodo.1219473}}

\acknowledgments

The research leading to these results has received funding from the European
Research Council under the European Union's Seventh Framework Program
(FP/2007--2013) / ERC grant Agreement No.~279954.  C.M.\ would like to thank
Antonio~J.~Cuesta, Bruce~A.~Bassett, Jarah~Evslin, and Hendrik~Hildebrandt for
their helpful comments, and to acknowledge the support from the joint research
training program of CAS and CNRS. We gratefully acknowledge the anonymous
reviewers for their commentary that helped improve this paper.

\software{CAMB\footnote{\url{http://camb.info/}} \citep{2000ApJ...538..473L,
cambzenodo}, matplotlib \citep{Hunter07, matplotlib222}, statsmodels
\citep{statsmodels}, NumPy, and SciPy.\footnote{\url{https://scipy.org/}}}

\appendix

\twocolumngrid
\section{Robustness of Gaussian Approximation for $A_B$ and $\eta$}
\label{sec:a:ab}

\begin{figure}
    \plotone{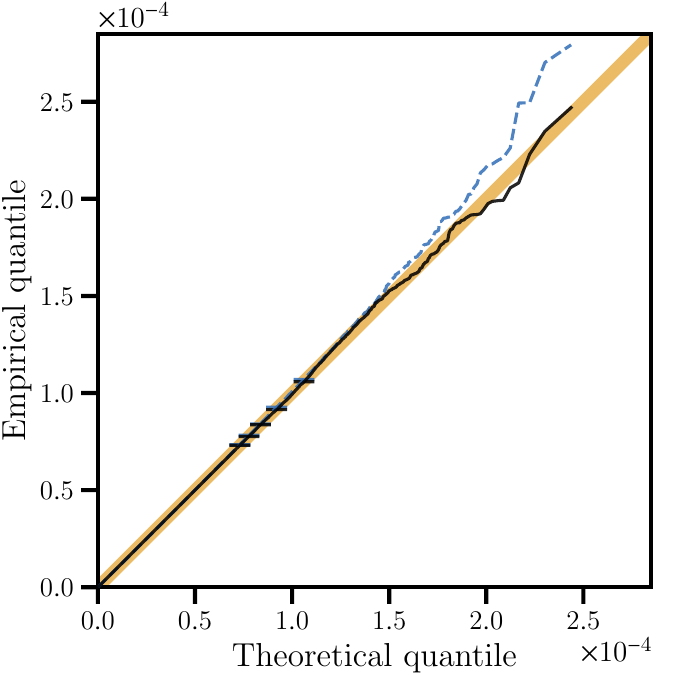}
    \caption{Q--Q plot of the squared Mahalanobis distance from the
	main $A_B$ samples versus the theoretical distribution assuming MVN.
	For readability, the line of equal distribution is displayed as an
	orange (shaded) band, and the data points (not shown) are connected by
	line segments.  Markers show the positions of the 95th--99th
	percentiles.  Results from SNIa + BOSS are shown in black (solid), and
	SNIa + WiggleZ in blue (dashed). \label{fig:a:qqmahal}} 
\end{figure}

We use the MC-estimated sample mean vectors and covariance matrices of $A_B$
obtained in Section~\ref{sec:ddsystematics} to approximate the \lcdm
main results by the multivariate normal (MVN) distribution and study the
robustness of the approximation with a graphical test.  If a random sample
with sample mean $\vv{\mu}$ and sample covariance $C$ is drawn from a
$d$-dimensional MVN distribution, it follows that the sample of the square of
the Mahalanobis distance from $\vv{\mu}$, defined as
\begin{equation}
    l^2(\vv{x}) = (\vv{x} - \vv{\mu})^{T} C^{-1} (\vv{x} - \vv{\mu}),
\end{equation}
has a beta distribution after scaling \citep{VK08}.  Namely,
\begin{equation}
    \label{eq:a:mahatheoretical}
    \frac{n l^2}{(n - 1)^2} \sim \text{Beta}\left( \frac{d}{2}, \frac{n - d -
    1}{2} \right).
\end{equation}
We plot the empirical quantiles of $n l^2 / (n - 1)^2$ against those of the
theoretical distribution in Figure~\ref{fig:a:qqmahal}.  Although both samples
show deviations from the theoretical distribution only noticeably after about
the 98th percentile, the tail distribution of the SNIa + BOSS sample behaves
better than the one from SNIa + WiggleZ that shows considerable deviation.

To understand these differences, we apply two complementary MVN tests, namely
the empirical characteristic function test \citep{HZ90} with bandwidth
parameter $\beta = 0.5$ and the sample skewness test \citep{Mardia70}.  It
should be noted that here we are not performing a hypothesis testing.  Indeed,
we already understand that MVN as a hypothesis is unlikely to be true: the
distribution form of $A_B$ is manifestly non-Gaussian, and its MC generation
is not an independent sampling process.  Instead, we employ the former test's
sensitivity to heavy tails and the latter's sensitivity to shape asymmetry to
explain the deviations in Figure~\ref{fig:a:qqmahal}.  The tests are applied
to random subsamples of size $n = 50$ and are repeated $10^4$ times.  The
rejection rates $r$ from the runs are compared with each other and with the
significance level parameter $\alpha = 0.05$.  For SNIa + BOSS,
\citeauthor{HZ90}'s test produces $r = 0.043$ and \citeauthor{Mardia70}'s
skewness test $r = 0.040$.  In contrast, for SNIa + WiggleZ both tests give $r
= 0.056$, in excess of $\alpha$.

The tests suggest the robustness of MVN approximation for $A_B$ from
SNIa + BOSS, but not for the one from SNIa + WiggleZ that displays greater
asymmetry and heavier tails.  Therefore, we omit the approximation for the
latter.

In future analysis, it may be necessary to factorize or invert the covariance
matrix.  For numerical stability, we increase the number of digits printed
here, as presented in Table~\ref{tab:a:meancov}.

\begin{deluxetable}{ccccc}
    \tablecolumns{5}
    \tablecaption{Sample Mean and Covariance of $A_B$ from SNIa +
	BOSS with the \textit{Planck} CP Set.\label{tab:a:meancov}}
    \tablehead{$z$ & \colhead{Mean} &
	\multicolumn{3}{c}{$\text{Covariance} \times 10^3$}}
    \startdata
    0.38 & 0.0952 & 4.176680 & 2.817193 & 2.791656 \\
    0.51 & 0.0514 & \nodata  & 4.259605 & 3.068827 \\
    0.61 & 0.0393 & \nodata  & \nodata  & 4.567452
    \enddata
    \tablecomments{To obtain the covariance matrix, multiply the values in the
    last three columns by $10^{-3}$.}
\end{deluxetable}

We calculate the probability $p$ of $A_B(z)$ being positive at all the three
redshifts given only the data, without model assumptions. This is the integral
of the data PDF over the infinite cell (octant) where all of the coordinates
are positive-valued.  Numerical quadrature using MVN approximation finds $p
\approx 0.64$.  Direct MC integration using the $A_B$ sample produces
essentially the same value, but its precision is limited by the MC sample
size.  We thus conclude that there is a slight preference of positive
extinction (Section~\ref{sec:ddsystematics}).

We also perform the three tests on the samples of $\eta$, and the results show
substantial deviation from MVN in all cases.

\bibliographystyle{aasjournal}
\bibliography{ms}

\begin{thebibliography}{}
\expandafter\ifx\csname natexlab\endcsname\relax\def\natexlab#1{#1}\fi
\providecommand{\url}[1]{\href{#1}{#1}}
\providecommand{\dodoi}[1]{doi:~\href{https://doi.org/#1}{\nolinkurl{#1}}}
\providecommand{\doeprint}[1]{\href{http://ascl.net/#1}{\nolinkurl{http://ascl.net/#1}}}
\providecommand{\doarXiv}[1]{\href{https://arxiv.org/abs/#1}{\nolinkurl{https://arxiv.org/abs/#1}}}

\bibitem[{{Abbott} {et~al.}(2017){Abbott}, {Abbott}, {Abbott}, {Acernese},
  {Ackley}, {Adams}, {Adams}, {Addesso}, {Adhikari}, {Adya}, \&
  et~al.}]{2017Natur.551...85A}
{Abbott}, B.~P., {Abbott}, R., {Abbott}, T.~D., {et~al.} 2017, Natur, 551, 85,
  \dodoi{10.1038/nature24471}

\bibitem[{{Alam} {et~al.}(2017){Alam}, {Ata}, {Bailey}, {Beutler}, {Bizyaev},
  {Blazek}, {Bolton}, {Brownstein}, {Burden}, {Chuang}, {Comparat}, {Cuesta},
  {Dawson}, {Eisenstein}, {Escoffier}, {Gil-Mar{\'{\i}}n}, {Grieb}, {Hand},
  {Ho}, {Kinemuchi}, {Kirkby}, {Kitaura}, {Malanushenko}, {Malanushenko},
  {Maraston}, {McBride}, {Nichol}, {Olmstead}, {Oravetz}, {Padmanabhan},
  {Palanque-Delabrouille}, {Pan}, {Pellejero-Ibanez}, {Percival}, {Petitjean},
  {Prada}, {Price-Whelan}, {Reid}, {Rodr{\'{\i}}guez-Torres}, {Roe}, {Ross},
  {Ross}, {Rossi}, {Rubi{\~n}o-Mart{\'{\i}}n}, {Saito}, {Salazar-Albornoz},
  {Samushia}, {S{\'a}nchez}, {Satpathy}, {Schlegel}, {Schneider},
  {Sc{\'o}ccola}, {Seo}, {Sheldon}, {Simmons}, {Slosar}, {Strauss}, {Swanson},
  {Thomas}, {Tinker}, {Tojeiro}, {Maga{\~n}a}, {Vazquez}, {Verde}, {Wake},
  {Wang}, {Weinberg}, {White}, {Wood-Vasey}, {Y{\`e}che}, {Zehavi}, {Zhai}, \&
  {Zhao}}]{2017MNRAS.470.2617A}
{Alam}, S., {Ata}, M., {Bailey}, S., {et~al.} 2017, \mnras, 470, 2617,
  \dodoi{10.1093/mnras/stx721}

\bibitem[{{Amanullah} {et~al.}(2010){Amanullah}, {Lidman}, {Rubin}, {Aldering},
  {Astier}, {Barbary}, {Burns}, {Conley}, {Dawson}, {Deustua}, {Doi}, {Fabbro},
  {Faccioli}, {Fakhouri}, {Folatelli}, {Fruchter}, {Furusawa}, {Garavini},
  {Goldhaber}, {Goobar}, {Groom}, {Hook}, {Howell}, {Kashikawa}, {Kim}, {Knop},
  {Kowalski}, {Linder}, {Meyers}, {Morokuma}, {Nobili}, {Nordin}, {Nugent},
  {{\"O}stman}, {Pain}, {Panagia}, {Perlmutter}, {Raux}, {Ruiz-Lapuente},
  {Spadafora}, {Strovink}, {Suzuki}, {Wang}, {Wood-Vasey}, {Yasuda}, \&
  {Supernova Cosmology Project}}]{2010ApJ...716..712A}
{Amanullah}, R., {Lidman}, C., {Rubin}, D., {et~al.} 2010, \apj, 716, 712,
  \dodoi{10.1088/0004-637X/716/1/712}

\bibitem[{{Anderson} {et~al.}(2014){Anderson}, {Aubourg}, {Bailey}, {Beutler},
  {Bhardwaj}, {Blanton}, {Bolton}, {Brinkmann}, {Brownstein}, {Burden},
  {Chuang}, {Cuesta}, {Dawson}, {Eisenstein}, {Escoffier}, {Gunn}, {Guo}, {Ho},
  {Honscheid}, {Howlett}, {Kirkby}, {Lupton}, {Manera}, {Maraston}, {McBride},
  {Mena}, {Montesano}, {Nichol}, {Nuza}, {Olmstead}, {Padmanabhan},
  {Palanque-Delabrouille}, {Parejko}, {Percival}, {Petitjean}, {Prada},
  {Price-Whelan}, {Reid}, {Roe}, {Ross}, {Ross}, {Sabiu}, {Saito}, {Samushia},
  {S{\'a}nchez}, {Schlegel}, {Schneider}, {Scoccola}, {Seo}, {Skibba},
  {Strauss}, {Swanson}, {Thomas}, {Tinker}, {Tojeiro}, {Maga{\~n}a}, {Verde},
  {Wake}, {Weaver}, {Weinberg}, {White}, {Xu}, {Y{\`e}che}, {Zehavi}, \&
  {Zhao}}]{2014MNRAS.441...24A}
{Anderson}, L., {Aubourg}, {\'E}., {Bailey}, S., {et~al.} 2014, \mnras, 441,
  24, \dodoi{10.1093/mnras/stu523}

\bibitem[{{Ballinger} {et~al.}(1996){Ballinger}, {Peacock}, \&
  {Heavens}}]{1996MNRAS.282..877B}
{Ballinger}, W.~E., {Peacock}, J.~A., \& {Heavens}, A.~F. 1996, \mnras, 282,
  877, \dodoi{10.1093/mnras/282.3.877}

\bibitem[{{Bassett} \& {Kunz}(2004)}]{2004PhRvD..69j1305B}
{Bassett}, B.~A., \& {Kunz}, M. 2004, \prd, 69, 101305,
  \dodoi{10.1103/PhysRevD.69.101305}

\bibitem[{{Betoule} {et~al.}(2014){Betoule}, {Kessler}, {Guy}, {Mosher},
  {Hardin}, {Biswas}, {Astier}, {El-Hage}, {Konig}, {Kuhlmann}, {Marriner},
  {Pain}, {Regnault}, {Balland}, {Bassett}, {Brown}, {Campbell}, {Carlberg},
  {Cellier-Holzem}, {Cinabro}, {Conley}, {D'Andrea}, {DePoy}, {Doi}, {Ellis},
  {Fabbro}, {Filippenko}, {Foley}, {Frieman}, {Fouchez}, {Galbany}, {Goobar},
  {Gupta}, {Hill}, {Hlozek}, {Hogan}, {Hook}, {Howell}, {Jha}, {Le Guillou},
  {Leloudas}, {Lidman}, {Marshall}, {M{\"o}ller}, {Mour{\~a}o}, {Neveu},
  {Nichol}, {Olmstead}, {Palanque-Delabrouille}, {Perlmutter}, {Prieto},
  {Pritchet}, {Richmond}, {Riess}, {Ruhlmann-Kleider}, {Sako}, {Schahmaneche},
  {Schneider}, {Smith}, {Sollerman}, {Sullivan}, {Walton}, \&
  {Wheeler}}]{2014AA...568A..22B}
{Betoule}, M., {Kessler}, R., {Guy}, J., {et~al.} 2014, \aap, 568, A22,
  \dodoi{10.1051/0004-6361/201423413}

\bibitem[{{Beutler} {et~al.}(2016){Beutler}, {Blake}, {Koda}, {Mar{\'{\i}}n},
  {Seo}, {Cuesta}, \& {Schneider}}]{2016MNRAS.455.3230B}
{Beutler}, F., {Blake}, C., {Koda}, J., {et~al.} 2016, \mnras, 455, 3230,
  \dodoi{10.1093/mnras/stv1943}

\bibitem[{{Blake} {et~al.}(2011{\natexlab{a}}){Blake}, {Davis}, {Poole},
  {Parkinson}, {Brough}, {Colless}, {Contreras}, {Couch}, {Croom},
  {Drinkwater}, {Forster}, {Gilbank}, {Gladders}, {Glazebrook}, {Jelliffe},
  {Jurek}, {Li}, {Madore}, {Martin}, {Pimbblet}, {Pracy}, {Sharp}, {Wisnioski},
  {Woods}, {Wyder}, \& {Yee}}]{2011MNRAS.415.2892B}
{Blake}, C., {Davis}, T., {Poole}, G.~B., {et~al.} 2011{\natexlab{a}}, \mnras,
  415, 2892, \dodoi{10.1111/j.1365-2966.2011.19077.x}

\bibitem[{{Blake} {et~al.}(2011{\natexlab{b}}){Blake}, {Glazebrook}, {Davis},
  {Brough}, {Colless}, {Contreras}, {Couch}, {Croom}, {Drinkwater}, {Forster},
  {Gilbank}, {Gladders}, {Jelliffe}, {Jurek}, {Li}, {Madore}, {Martin},
  {Pimbblet}, {Poole}, {Pracy}, {Sharp}, {Wisnioski}, {Woods}, {Wyder}, \&
  {Yee}}]{2011MNRAS.418.1725B}
{Blake}, C., {Glazebrook}, K., {Davis}, T.~M., {et~al.} 2011{\natexlab{b}},
  \mnras, 418, 1725, \dodoi{10.1111/j.1365-2966.2011.19606.x}

\bibitem[{{Blake} {et~al.}(2012){Blake}, {Brough}, {Colless}, {Contreras},
  {Couch}, {Croom}, {Croton}, {Davis}, {Drinkwater}, {Forster}, {Gilbank},
  {Gladders}, {Glazebrook}, {Jelliffe}, {Jurek}, {Li}, {Madore}, {Martin},
  {Pimbblet}, {Poole}, {Pracy}, {Sharp}, {Wisnioski}, {Woods}, {Wyder}, \&
  {Yee}}]{2012MNRAS.425..405B}
{Blake}, C., {Brough}, S., {Colless}, M., {et~al.} 2012, \mnras, 425, 405,
  \dodoi{10.1111/j.1365-2966.2012.21473.x}

\bibitem[{{Bocquet} {et~al.}(2015){Bocquet}, {Saro}, {Mohr}, {Aird}, {Ashby},
  {Bautz}, {Bayliss}, {Bazin}, {Benson}, {Bleem}, {Brodwin}, {Carlstrom},
  {Chang}, {Chiu}, {Cho}, {Clocchiatti}, {Crawford}, {Crites}, {Desai}, {de
  Haan}, {Dietrich}, {Dobbs}, {Foley}, {Forman}, {Gangkofner}, {George},
  {Gladders}, {Gonzalez}, {Halverson}, {Hennig}, {Hlavacek-Larrondo}, {Holder},
  {Holzapfel}, {Hrubes}, {Jones}, {Keisler}, {Knox}, {Lee}, {Leitch}, {Liu},
  {Lueker}, {Luong-Van}, {Marrone}, {McDonald}, {McMahon}, {Meyer}, {Mocanu},
  {Murray}, {Padin}, {Pryke}, {Reichardt}, {Rest}, {Ruel}, {Ruhl},
  {Saliwanchik}, {Sayre}, {Schaffer}, {Shirokoff}, {Spieler}, {Stalder},
  {Stanford}, {Staniszewski}, {Stark}, {Story}, {Stubbs}, {Vanderlinde},
  {Vieira}, {Vikhlinin}, {Williamson}, {Zahn}, \&
  {Zenteno}}]{2015ApJ...799..214B}
{Bocquet}, S., {Saro}, A., {Mohr}, J.~J., {et~al.} 2015, \apj, 799, 214,
  \dodoi{10.1088/0004-637X/799/2/214}

\bibitem[{{Bonamente} {et~al.}(2006){Bonamente}, {Joy}, {LaRoque}, {Carlstrom},
  {Reese}, \& {Dawson}}]{2006ApJ...647...25B}
{Bonamente}, M., {Joy}, M.~K., {LaRoque}, S.~J., {et~al.} 2006, \apj, 647, 25,
  \dodoi{10.1086/505291}

\bibitem[{{Cardone} {et~al.}(2012){Cardone}, {Spiro}, {Hook}, \&
  {Scaramella}}]{2012PhRvD..85l3510C}
{Cardone}, V.~F., {Spiro}, S., {Hook}, I., \& {Scaramella}, R. 2012, \prd, 85,
  123510, \dodoi{10.1103/PhysRevD.85.123510}

\bibitem[{{Chiu} {et~al.}(2016){Chiu}, {Mohr}, {McDonald}, {Bocquet}, {Ashby},
  {Bayliss}, {Benson}, {Bleem}, {Brodwin}, {Desai}, {Dietrich}, {Forman},
  {Gangkofner}, {Gonzalez}, {Hennig}, {Liu}, {Reichardt}, {Saro}, {Stalder},
  {Stanford}, {Song}, {Schrabback}, {{\v S}uhada}, {Strazzullo}, \&
  {Zenteno}}]{2016MNRAS.455..258C}
{Chiu}, I., {Mohr}, J., {McDonald}, M., {et~al.} 2016, \mnras, 455, 258,
  \dodoi{10.1093/mnras/stv2303}

\bibitem[{{Chluba}(2014)}]{2014MNRAS.443.1881C}
{Chluba}, J. 2014, \mnras, 443, 1881, \dodoi{10.1093/mnras/stu1260}

\bibitem[{{Corasaniti}(2006)}]{2006MNRAS.372..191C}
{Corasaniti}, P.-S. 2006, \mnras, 372, 191,
  \dodoi{10.1111/j.1365-2966.2006.10825.x}

\bibitem[{{Crocce} \& {Scoccimarro}(2008)}]{2008PhRvD..77b3533C}
{Crocce}, M., \& {Scoccimarro}, R. 2008, \prd, 77, 023533,
  \dodoi{10.1103/PhysRevD.77.023533}

\bibitem[{{Cyburt} {et~al.}(2016){Cyburt}, {Fields}, {Olive}, \&
  {Yeh}}]{2016RvMP...88a5004C}
{Cyburt}, R.~H., {Fields}, B.~D., {Olive}, K.~A., \& {Yeh}, T.-H. 2016, RvMP,
  88, 015004, \dodoi{10.1103/RevModPhys.88.015004}

\bibitem[{{De Bernardis} {et~al.}(2006){De Bernardis}, {Giusarma}, \&
  {Melchiorri}}]{2006IJMPD..15..759D}
{De Bernardis}, F., {Giusarma}, E., \& {Melchiorri}, A. 2006, IJMPD, 15, 759,
  \dodoi{10.1142/S0218271806008486}

\bibitem[{Droettboom {et~al.}(2018)Droettboom, Hunter, Caswell, Firing,
  Nielsen, Lee, {Sales de Andrade}, Varoquaux, Root, Stansby, \&
  et~al.}]{matplotlib222}
Droettboom, M., Hunter, J., Caswell, T.~A., {et~al.} 2018, matplotlib, 2.2.2,
  Zenodo, \dodoi{10.5281/zenodo.1202077}.
\newblock \url{https://doi.org/10.5281/zenodo.1202077}

\bibitem[{{Efstathiou}(2014)}]{2014MNRAS.440.1138E}
{Efstathiou}, G. 2014, \mnras, 440, 1138, \dodoi{10.1093/mnras/stu278}

\bibitem[{{Eisenstein} {et~al.}(2005){Eisenstein}, {Zehavi}, {Hogg},
  {Scoccimarro}, {Blanton}, {Nichol}, {Scranton}, {Seo}, {Tegmark}, {Zheng},
  {Anderson}, {Annis}, {Bahcall}, {Brinkmann}, {Burles}, {Castander},
  {Connolly}, {Csabai}, {Doi}, {Fukugita}, {Frieman}, {Glazebrook}, {Gunn},
  {Hendry}, {Hennessy}, {Ivezi{\'c}}, {Kent}, {Knapp}, {Lin}, {Loh}, {Lupton},
  {Margon}, {McKay}, {Meiksin}, {Munn}, {Pope}, {Richmond}, {Schlegel},
  {Schneider}, {Shimasaku}, {Stoughton}, {Strauss}, {SubbaRao}, {Szalay},
  {Szapudi}, {Tucker}, {Yanny}, \& {York}}]{2005ApJ...633..560E}
{Eisenstein}, D.~J., {Zehavi}, I., {Hogg}, D.~W., {et~al.} 2005, \apj, 633,
  560, \dodoi{10.1086/466512}

\bibitem[{{Ellis}(1971)}]{E71}
{Ellis}, G.~F.~R. 1971, in {Proceedings of the International School of Physics
  ``Enrico Fermi'', Course 47: General Relativity and Cosmology}, ed. R.~K.
  {Sachs} (New York, US: {Academic Press}), 104--182

\bibitem[{{Ellis}(2009)}]{2009GReGr..41..581E}
{Ellis}, G.~F.~R. 2009, GReGr, 41, 581, \dodoi{10.1007/s10714-009-0760-7}

\bibitem[{{Etherington}(1933)}]{E33}
{Etherington}, I.~M.~H. 1933, PMag, 7, 761

\bibitem[{{Etherington}(2007)}]{2007GReGr..39.1055E}
---. 2007, GReGr, 39, 1055, \dodoi{10.1007/s10714-007-0447-x}

\bibitem[{{Euclid Science Study Team}(2010)}]{EuclidScird}
{Euclid Science Study Team}. 2010, {Euclid Science Requirements Document},
  Tech. Rep. DEM-SA-Dc-00001, {European Space Research and Technology Centre,
  ESA}, Noordwijk, NL

\bibitem[{{Evslin}(2016)}]{2016PDU....14...57E}
{Evslin}, J. 2016, PDU, 14, 57, \dodoi{10.1016/j.dark.2016.09.005}

\bibitem[{{Flegal} {et~al.}(2008){Flegal}, {Haran}, \& {Jones}}]{f08}
{Flegal}, J.~M., {Haran}, M., \& {Jones}, G.~L. 2008, StatSc, 23, 250,
  \dodoi{10.1214/08-STS257}

\bibitem[{{Fu} \& {Li}(2017)}]{2017IJMPD..2650097F}
{Fu}, X., \& {Li}, P. 2017, IJMPD, 26, 1750097,
  \dodoi{10.1142/S0218271817500973}

\bibitem[{{Gonzalez} {et~al.}(2013){Gonzalez}, {Sivanandam}, {Zabludoff}, \&
  {Zaritsky}}]{2013ApJ...778...14G}
{Gonzalez}, A.~H., {Sivanandam}, S., {Zabludoff}, A.~I., \& {Zaritsky}, D.
  2013, \apj, 778, 14, \dodoi{10.1088/0004-637X/778/1/14}

\bibitem[{{Goobar} \& {Leibundgut}(2011)}]{2011ARNPS..61..251G}
{Goobar}, A., \& {Leibundgut}, B. 2011, ARNPS, 61, 251,
  \dodoi{10.1146/annurev-nucl-102010-130434}

\bibitem[{{Grandis} {et~al.}(2016){Grandis}, {Rapetti}, {Saro}, {Mohr}, \&
  {Dietrich}}]{2016MNRAS.463.1416G}
{Grandis}, S., {Rapetti}, D., {Saro}, A., {Mohr}, J.~J., \& {Dietrich}, J.~P.
  2016, \mnras, 463, 1416, \dodoi{10.1093/mnras/stw2028}

\bibitem[{{Henze} \& {Zirkler}(1990)}]{HZ90}
{Henze}, N., \& {Zirkler}, B. 1990, Commun.\ Stat.,\ Theory Methods, 19, 3595,
  \dodoi{10.1080/03610929008830400}

\bibitem[{{Hildebrandt} {et~al.}(2017){Hildebrandt}, {Viola}, {Heymans},
  {Joudaki}, {Kuijken}, {Blake}, {Erben}, {Joachimi}, {Klaes}, {Miller},
  {Morrison}, {Nakajima}, {Verdoes Kleijn}, {Amon}, {Choi}, {Covone}, {de
  Jong}, {Dvornik}, {Fenech Conti}, {Grado}, {Harnois-D{\'e}raps}, {Herbonnet},
  {Hoekstra}, {K{\"o}hlinger}, {McFarland}, {Mead}, {Merten}, {Napolitano},
  {Peacock}, {Radovich}, {Schneider}, {Simon}, {Valentijn}, {van den Busch},
  {van Uitert}, \& {Van Waerbeke}}]{2017MNRAS.465.1454H}
{Hildebrandt}, H., {Viola}, M., {Heymans}, C., {et~al.} 2017, \mnras, 465,
  1454, \dodoi{10.1093/mnras/stw2805}

\bibitem[{{Holanda} {et~al.}(2016){Holanda}, {Busti}, \&
  {Alcaniz}}]{2016JCAP...02..054H}
{Holanda}, R.~F.~L., {Busti}, V.~C., \& {Alcaniz}, J.~S. 2016, \jcap, 2, 054,
  \dodoi{10.1088/1475-7516/2016/02/054}

\bibitem[{{Holanda} {et~al.}(2017){Holanda}, {Busti}, {Lima}, \&
  {Alcaniz}}]{2017JCAP...09..039H}
{Holanda}, R.~F.~L., {Busti}, V.~C., {Lima}, F.~S., \& {Alcaniz}, J.~S. 2017,
  \jcap, 9, 039, \dodoi{10.1088/1475-7516/2017/09/039}

\bibitem[{{Holanda} {et~al.}(2010){Holanda}, {Lima}, \&
  {Ribeiro}}]{2010ApJ...722L.233H}
{Holanda}, R.~F.~L., {Lima}, J.~A.~S., \& {Ribeiro}, M.~B. 2010, \apjl, 722,
  L233, \dodoi{10.1088/2041-8205/722/2/L233}

\bibitem[{{Hu} {et~al.}(1995){Hu}, {Scott}, {Sugiyama}, \&
  {White}}]{1995PhRvD..52.5498H}
{Hu}, W., {Scott}, D., {Sugiyama}, N., \& {White}, M. 1995, \prd, 52, 5498,
  \dodoi{10.1103/PhysRevD.52.5498}

\bibitem[{{Hunter}(2007)}]{Hunter07}
{Hunter}, J.~D. 2007, CSE, 9, 90, \dodoi{10.1109/MCSE.2007.55}

\bibitem[{{Lewis} {et~al.}(2000){Lewis}, {Challinor}, \&
  {Lasenby}}]{2000ApJ...538..473L}
{Lewis}, A., {Challinor}, A., \& {Lasenby}, A. 2000, \apj, 538, 473,
  \dodoi{10.1086/309179}

\bibitem[{Lewis {et~al.}(2017)Lewis, Mead, Vehreschild, Millea, Bird, Casarini,
  \& Torrado}]{cambzenodo}
Lewis, A., Mead, A., Vehreschild, A., {et~al.} 2017, CAMB, {August 2017},
  \dodoi{10.5281/zenodo.844843}

\bibitem[{{Liang} {et~al.}(2013){Liang}, {Li}, {Wu}, {Cao}, {Liao}, \&
  {Zhu}}]{2013MNRAS.436.1017L}
{Liang}, N., {Li}, Z., {Wu}, P., {et~al.} 2013, \mnras, 436, 1017,
  \dodoi{10.1093/mnras/stt1589}

\bibitem[{{Liao} {et~al.}(2015){Liao}, {Avgoustidis}, \&
  {Li}}]{2015PhRvD..92l3539L}
{Liao}, K., {Avgoustidis}, A., \& {Li}, Z. 2015, \prd, 92, 123539,
  \dodoi{10.1103/PhysRevD.92.123539}

\bibitem[{{Liao} {et~al.}(2016){Liao}, {Li}, {Cao}, {Biesiada}, {Zheng}, \&
  {Zhu}}]{2016ApJ...822...74L}
{Liao}, K., {Li}, Z., {Cao}, S., {et~al.} 2016, \apj, 822, 74,
  \dodoi{10.3847/0004-637X/822/2/74}

\bibitem[{{LSST Science Collaboration}(2009)}]{2009arXiv0912.0201L}
{LSST Science Collaboration}. 2009, preprint.
\newblock \doarXiv{0912.0201}

\bibitem[{{Ma} {et~al.}(2016){Ma}, {Corasaniti}, \&
  {Bassett}}]{2016MNRAS.463.1651M}
{Ma}, C., {Corasaniti}, P.-S., \& {Bassett}, B.~A. 2016, \mnras, 463, 1651,
  \dodoi{10.1093/mnras/stw2069}

\bibitem[{{Mardia}(1970)}]{Mardia70}
{Mardia}, K.~V. 1970, Biometrika, 57, 519, \dodoi{10.1093/biomet/57.3.519}

\bibitem[{{Max-Moerbeck} {et~al.}(2014){Max-Moerbeck}, {Richards}, {Hovatta},
  {Pavlidou}, {Pearson}, \& {Readhead}}]{2014MNRAS.445..437M}
{Max-Moerbeck}, W., {Richards}, J.~L., {Hovatta}, T., {et~al.} 2014, \mnras,
  445, 437, \dodoi{10.1093/mnras/stu1707}

\bibitem[{{Mehta} {et~al.}(2012){Mehta}, {Cuesta}, {Xu}, {Eisenstein}, \&
  {Padmanabhan}}]{2012MNRAS.427.2168M}
{Mehta}, K.~T., {Cuesta}, A.~J., {Xu}, X., {Eisenstein}, D.~J., \&
  {Padmanabhan}, N. 2012, \mnras, 427, 2168,
  \dodoi{10.1111/j.1365-2966.2012.21112.x}

\bibitem[{{M{\'e}nard} {et~al.}(2010{\natexlab{a}}){M{\'e}nard}, {Kilbinger},
  \& {Scranton}}]{2010MNRAS.406.1815M}
{M{\'e}nard}, B., {Kilbinger}, M., \& {Scranton}, R. 2010{\natexlab{a}},
  \mnras, 406, 1815, \dodoi{10.1111/j.1365-2966.2010.16464.x}

\bibitem[{{M{\'e}nard} {et~al.}(2010{\natexlab{b}}){M{\'e}nard}, {Scranton},
  {Fukugita}, \& {Richards}}]{2010MNRAS.405.1025M}
{M{\'e}nard}, B., {Scranton}, R., {Fukugita}, M., \& {Richards}, G.
  2010{\natexlab{b}}, \mnras, 405, 1025,
  \dodoi{10.1111/j.1365-2966.2010.16486.x}

\bibitem[{{Meng} {et~al.}(2012){Meng}, {Zhang}, {Zhan}, \&
  {Wang}}]{2012ApJ...745...98M}
{Meng}, X.-L., {Zhang}, T.-J., {Zhan}, H., \& {Wang}, X. 2012, \apj, 745, 98,
  \dodoi{10.1088/0004-637X/745/1/98}

\bibitem[{{More} {et~al.}(2009){More}, {Bovy}, \& {Hogg}}]{2009ApJ...696.1727M}
{More}, S., {Bovy}, J., \& {Hogg}, D.~W. 2009, \apj, 696, 1727,
  \dodoi{10.1088/0004-637X/696/2/1727}

\bibitem[{{Nair} {et~al.}(2012){Nair}, {Jhingan}, \&
  {Jain}}]{2012JCAP...12..028N}
{Nair}, R., {Jhingan}, S., \& {Jain}, D. 2012, \jcap, 12, 028,
  \dodoi{10.1088/1475-7516/2012/12/028}

\bibitem[{{Planck Collaboration}(2016{\natexlab{a}})}]{2016AA...594A..13P}
{Planck Collaboration}. 2016{\natexlab{a}}, \aap, 594, A13,
  \dodoi{10.1051/0004-6361/201525830}

\bibitem[{{Planck Collaboration}(2016{\natexlab{b}})}]{2016AA...594A..14P}
---. 2016{\natexlab{b}}, \aap, 594, A14, \dodoi{10.1051/0004-6361/201525814}

\bibitem[{{Rana} {et~al.}(2016){Rana}, {Jain}, {Mahajan}, \&
  {Mukherjee}}]{2016JCAP...07..026R}
{Rana}, A., {Jain}, D., {Mahajan}, S., \& {Mukherjee}, A. 2016, \jcap, 7, 026,
  \dodoi{10.1088/1475-7516/2016/07/026}

\bibitem[{{Rana} {et~al.}(2017){Rana}, {Jain}, {Mahajan}, {Mukherjee}, \&
  {Holanda}}]{2017JCAP...07..010R}
{Rana}, A., {Jain}, D., {Mahajan}, S., {Mukherjee}, A., \& {Holanda}, R.~F.~L.
  2017, \jcap, 7, 010, \dodoi{10.1088/1475-7516/2017/07/010}

\bibitem[{{R{\"a}s{\"a}nen} {et~al.}(2016){R{\"a}s{\"a}nen}, {V{\"a}liviita},
  \& {Kosonen}}]{2016JCAP...04..050R}
{R{\"a}s{\"a}nen}, S., {V{\"a}liviita}, J., \& {Kosonen}, V. 2016, \jcap, 4,
  050, \dodoi{10.1088/1475-7516/2016/04/050}

\bibitem[{{Rasera} {et~al.}(2014){Rasera}, {Corasaniti}, {Alimi}, {Bouillot},
  {Reverdy}, \& {Balm{\`e}s}}]{2014MNRAS.440.1420R}
{Rasera}, Y., {Corasaniti}, P.-S., {Alimi}, J.-M., {et~al.} 2014, \mnras, 440,
  1420, \dodoi{10.1093/mnras/stu295}

\bibitem[{{Riess} {et~al.}(2016){Riess}, {Macri}, {Hoffmann}, {Scolnic},
  {Casertano}, {Filippenko}, {Tucker}, {Reid}, {Jones}, {Silverman},
  {Chornock}, {Challis}, {Yuan}, {Brown}, \& {Foley}}]{2016ApJ...826...56R}
{Riess}, A.~G., {Macri}, L.~M., {Hoffmann}, S.~L., {et~al.} 2016, \apj, 826,
  56, \dodoi{10.3847/0004-637X/826/1/56}

\bibitem[{{Riess} {et~al.}(2018){Riess}, {Casertano}, {Yuan}, {Macri},
  {Anderson}, {MacKenty}, {Bowers}, {Clubb}, {Filippenko}, {Jones}, \&
  {Tucker}}]{2018ApJ...855..136R}
{Riess}, A.~G., {Casertano}, S., {Yuan}, W., {et~al.} 2018, \apj, 855, 136,
  \dodoi{10.3847/1538-4357/aaadb7}

\bibitem[{{Rigault} {et~al.}(2015){Rigault}, {Aldering}, {Kowalski}, {Copin},
  {Antilogus}, {Aragon}, {Bailey}, {Baltay}, {Baugh}, {Bongard}, {Boone},
  {Buton}, {Chen}, {Chotard}, {Fakhouri}, {Feindt}, {Fagrelius}, {Fleury},
  {Fouchez}, {Gangler}, {Hayden}, {Kim}, {Leget}, {Lombardo}, {Nordin}, {Pain},
  {Pecontal}, {Pereira}, {Perlmutter}, {Rabinowitz}, {Runge}, {Rubin},
  {Saunders}, {Smadja}, {Sofiatti}, {Suzuki}, {Tao}, \&
  {Weaver}}]{2015ApJ...802...20R}
{Rigault}, M., {Aldering}, G., {Kowalski}, M., {et~al.} 2015, \apj, 802, 20,
  \dodoi{10.1088/0004-637X/802/1/20}

\bibitem[{{Santos-da-Costa} {et~al.}(2015){Santos-da-Costa}, {Busti}, \&
  {Holanda}}]{2015JCAP...10..061S}
{Santos-da-Costa}, S., {Busti}, V.~C., \& {Holanda}, R.~F.~L. 2015, \jcap, 10,
  061, \dodoi{10.1088/1475-7516/2015/10/061}

\bibitem[{{Schneider} \& {Sluse}(2013)}]{2013AA...559A..37S}
{Schneider}, P., \& {Sluse}, D. 2013, \aap, 559, A37,
  \dodoi{10.1051/0004-6361/201321882}

\bibitem[{Seabold {et~al.}(2017)Seabold, Perktold, Fulton, Shedden, Grana,
  Sheppard, Arel-Bundock, McKinney, Langmore, Baker, Gommers, Giampieri, Brett,
  Hobson, Millman, \& et~al.}]{statsmodels}
Seabold, S., Perktold, J., Fulton, C., {et~al.} 2017, statsmodels, 0.8.0,
  Zenodo, \dodoi{10.5281/zenodo.275519}

\bibitem[{{Seager} {et~al.}(2000){Seager}, {Sasselov}, \&
  {Scott}}]{2000ApJS..128..407S}
{Seager}, S., {Sasselov}, D.~D., \& {Scott}, D. 2000, \apjs, 128, 407,
  \dodoi{10.1086/313388}

\bibitem[{{Uzan} {et~al.}(2004){Uzan}, {Aghanim}, \&
  {Mellier}}]{2004PhRvD..70h3533U}
{Uzan}, J.-P., {Aghanim}, N., \& {Mellier}, Y. 2004, \prd, 70, 083533,
  \dodoi{10.1103/PhysRevD.70.083533}

\bibitem[{{Ververidis} \& {Kotropoulos}(2008)}]{VK08}
{Ververidis}, D., \& {Kotropoulos}, C. 2008, ITSP, 56, 2797,
  \dodoi{10.1109/TSP.2008.917350}

\bibitem[{{Wu} {et~al.}(2015){Wu}, {Li}, {Liu}, \& {Yu}}]{2015PhRvD..92b3520W}
{Wu}, P., {Li}, Z., {Liu}, X., \& {Yu}, H. 2015, \prd, 92, 023520,
  \dodoi{10.1103/PhysRevD.92.023520}

\bibitem[{{Yang} {et~al.}(2013){Yang}, {Yu}, {Zhang}, \&
  {Zhang}}]{2013ApJ...777L..24Y}
{Yang}, X., {Yu}, H.-R., {Zhang}, Z.-S., \& {Zhang}, T.-J. 2013, \apjl, 777,
  L24, \dodoi{10.1088/2041-8205/777/2/L24}

\end{thebibliography}

\end{document}